\documentclass{article}

\usepackage{fullpage} 
\usepackage{amssymb} 
\usepackage[all,dvips]{xy} 
\usepackage{url}
\usepackage{fullpage}
\usepackage{rotating,stmaryrd}

\usepackage{theorem}
\newtheorem{theo}{Theorem}[section]
\newtheorem{prop}[theo]{Proposition}
\newtheorem{coro}[theo]{Corollary}
\newtheorem{lemm}[theo]{Lemma}
\theorembodyfont{\normalfont}
\newtheorem{defi}[theo]{Definition}
\newtheorem{exam}[theo]{Example}
\newtheorem{rema}[theo]{Remark}

\newenvironment{proof}{\textit{Proof.} }{$\Box$}

\newcommand{\ol}{\overline}  
 
\newcommand{\eps}{\varepsilon} 
\newcommand{\cosl}{\!\!\downarrow\!\!} 
\newcommand{\tuple}[1]{\langle#1\rangle} 
\newcommand{\tu}{\langle\,\rangle} 
\newcommand{\curry}[1]{\ulcorner\!{#1}\!\urcorner} 
\newcommand{\kl}[1]{[#1]} 
\newcommand{\PO}{{\scriptstyle [P.O.]}} 
\newcommand{\stimes}{\!\times\!} 
\newcommand{\md}{*}  
\newcommand{\logL}{\mathcal{L}} 
\newcommand{\frsyn}[2]{{#1}/{#2}}
\newcommand{\fr}[2]{{#2}\backslash{#1}}
\newcommand{\entail}{$\!\!\!\!\!\! \xymatrix@C=3pc{
    \mbox{} \ar[r]_{\subseteq} & \mbox{} \ar@<-1ex>@{-->}[l]  \\}  \!\!\!\!\!\!$}
\newcommand{\entailhigh}{$\!\!\!\!\!\! \xymatrix@C=3pc@R=0pc{ \\
    \mbox{} \ar[r]_{\subseteq} & \mbox{} \ar@<-1ex>@{-->}[l]  \\}  \!\!\!\!\!\!$}

\newcommand{\hsp}{\hspace{5mm}}  
\newcommand{\haut}{\rule[-10pt]{0pt}{25pt}}
\newcommand{\hau}{\rule[-5pt]{0pt}{0pt}}

\newcommand{\col}[1]{\multicolumn{1}{c}{#1}}
\newcommand{\cols}[2]{\multicolumn{#1}{|c|}{#2}}

\newcommand{\isoto}{\stackrel{\simeq}{\rightarrow}} 
\newcommand{\iso}{\cong} 
\newcommand{\To}{\Rightarrow} 
\newcommand{\trnat}{\begin{turn}{-20}\ensuremath{\Uparrow}\end{turn}} 
\newcommand{\upseteq}{\begin{turn}{90}\ensuremath{\subseteq}\end{turn}} 
\newcommand{\rpto}{\rightsquigarrow} 
\newcommand{\rupto}[1]{\stackrel{#1}{\longrightarrow}}
\newcommand{\lupto}[1]{\stackrel{#1}{\longleftarrow}}

\newcommand{\bN}{\mathbb{N}} 
\newcommand{\bZ}{\mathbb{Z}} 
\newcommand{\bL}{\mathbb{L}} 
\newcommand{\bA}{\mathbb{A}} 
\newcommand{\setX}{\mathbb{X}}  
\newcommand{\setY}{\mathbb{Y}}   

\newcommand{\uno}{1}  

\newcommand{\catC}{\mathbf{C}}   
\newcommand{\catS}{\mathbf{S}}   
\newcommand{\catT}{\mathbf{T}}   
\newcommand{\Hom}{\mathit{Hom}} 
\newcommand{\eq}{\mathit{eq}} 
\newcommand{\dec}{\mathit{dec}} 

\newcommand{\funY}{\mathcal{Y}} 

\newcommand{\Spec}{\mathrm{Sp}} 
\newcommand{\Tt}{\Theta} 
\renewcommand{\tt}{\theta} 
\newcommand{\Ss}{\Sigma} 
\renewcommand{\ss}{\sigma} 
\newcommand{\Set}{\mathit{Set}} 
\newcommand{\HH}{\mathcal{H}} 
\newcommand{\CC}{\mathcal{C}} 
\newcommand{\Mod}{\mathit{Mod}} 
\newcommand{\catD}{\mathit{\Delta}} 
\newcommand{\sgp}{\mathit{sgp}} 
\newcommand{\mgm}{\mathit{mgm}} 
\newcommand{\mon}{\mathit{mon}} 
\newcommand{\dm}{\mathit{dm}} 
\newcommand{\nat}{\mathit{nat}} 
\newcommand{\st}{\mathit{st}} 
\newcommand{\oper}{\mathit{op}} 

\newcommand{\id}{\mathit{id}} 
\newcommand{\proj}{\mathit{pr}}  
\newcommand{\prd}{\mathit{prd}} 
\newcommand{\unt}{e} 
\newcommand{\dif}{\mathit{dif}} 
\newcommand{\lup}{\mathit{lookup}} 
\newcommand{\upd}{\mathit{update}} 

\newcommand{\skE}{\mathbf{E}}   
\newcommand{\olskE}{\ol{\skE}}  
\newcommand{\ske}{\mathbf{e}}   
\newcommand{\Rea}{\mathit{Real}} 
\newcommand{\gr}{\mathit{gr}} 
\newcommand{\cat}{\mathit{cat}} 
\newcommand{\thry}{\mathit{th}} 
\newcommand{\spec}{\mathit{sp}} 
\newcommand{\grco}{\mathit{gr\_comp}} 
\newcommand{\grid}{\mathit{gr\_id}} 
\newcommand{\greq}{\mathit{gr\_eq}} 
\newcommand{\grpr}{\mathit{gr\_prod}} 
\newcommand{\grpair}{\mathit{gr\_pair}} 
\newcommand{\grfi}{\mathit{gr\_fin}} 
\newcommand{\grcoll}{\mathit{gr\_coll}} 
\newcommand{\undec}{\mathit{undec}} 
\newcommand{\expan}{\mathit{exp}} 

\newcommand{\Type}{\mathtt{Type}}
\newcommand{\Term}{\mathtt{Term}}
\newcommand{\dom}{\mathtt{dom}}
\newcommand{\codom}{\mathtt{codom}}
\newcommand{\selid}{\mathtt{selid}}
\newcommand{\Selid}{\mathtt{Selid}}
\newcommand{\Cons}{\mathtt{Cons}}
\newcommand{\fst}{\mathtt{fst}}
\newcommand{\snd}{\mathtt{snd}}
\newcommand{\midd}{\mathtt{middle}}
\newcommand{\comp}{\mathtt{comp}}
\newcommand{\Comp}{\mathtt{Comp}}
\newcommand{\prbone}{\mathtt{b1}}
\newcommand{\prbtwo}{\mathtt{b2}}
\newcommand{\prbn}{\mathtt{bn}}
\newcommand{\nType}{\mathtt{Type}^\mathtt{n}}
\newcommand{\zType}{\mathtt{Type}^\mathtt{0}}
\newcommand{\bType}{\mathtt{Type}^\mathtt{2}}
\newcommand{\nCone}{\mathtt{n\texttt{-}Cone}}
\newcommand{\zCone}{\mathtt{0\texttt{-}Cone}}
\newcommand{\bCone}{\mathtt{2\texttt{-}Cone}}
\newcommand{\prcone}{\mathtt{c1}}
\newcommand{\prctwo}{\mathtt{c2}}
\newcommand{\prcn}{\mathtt{cn}}
\newcommand{\vertex}{\mathtt{vertex}}
\newcommand{\nprod}{\mathtt{n\texttt{-}prod}}
\newcommand{\bprod}{\mathtt{2\texttt{-}prod}}
\newcommand{\fin}{\mathtt{final}}
\newcommand{\nProd}{\mathtt{n\texttt{-}Prod}}
\newcommand{\bProd}{\mathtt{2\texttt{-}Prod}}
\newcommand{\Fin}{\mathtt{Final}}
\newcommand{\nbase}{\mathtt{n\texttt{-}base}} 
\newcommand{\bbase}{\mathtt{2\texttt{-}base}} 
\newcommand{\ntuple}{\mathtt{n\texttt{-}tuple}} 
\newcommand{\btuple}{\mathtt{2\texttt{-}tuple}} 
\newcommand{\nTuple}{\mathtt{n\texttt{-}Tuple}} 
\newcommand{\bTuple}{\mathtt{2\texttt{-}Tuple}} 
\newcommand{\ttA}{\mathtt{A}} 
 
\newcommand{\Unit}{\mathtt{Unit}}
\newcommand{\ttt}{\mathtt{t}}
\newcommand{\ttid}{\mathtt{id}}
\newcommand{\ttE}{\mathtt{E}}
\newcommand{\tte}{\mathtt{e}}
\newcommand{\deco}{\mathtt{.x}} 
\newcommand{\pur}{\mathtt{.p}} 
\newcommand{\gen}{\mathtt{.g}} 
\newcommand{\Refl}{\mathtt{Refl}} 
\newcommand{\tti}{\mathtt{i}}
\newcommand{\ttiz}{\mathtt{i0}}
\newcommand{\ttj}{\mathtt{j}}
\newcommand{\ttjz}{\mathtt{j0}}
\newcommand{\Para}{\mathtt{Para}}
\newcommand{\rght}{\mathtt{right}}
\newcommand{\lft}{\mathtt{left}}
\newcommand{\Equa}{\mathtt{Equa}}
\newcommand{\equa}{\mathtt{equa}}
\newcommand{\zbase}{\mathtt{0\texttt{-}base}}
\newcommand{\zprod}{\mathtt{0\texttt{-}prod}}
\newcommand{\zProd}{\mathtt{0\texttt{-}Prod}}

\newcommand{\Pair}{\mathtt{Pair}}
\newcommand{\pair}{\mathtt{pair}}
\newcommand{\bdom}{\mathtt{2\texttt{-}dom}} 
\newcommand{\bcodom}{\mathtt{2\texttt{-}codom}} 
\newcommand{\Coll}{\mathtt{Coll}}
\newcommand{\coll}{\mathtt{coll}}
\newcommand{\zdom}{\mathtt{0\texttt{-}dom}} 
\newcommand{\zcodom}{\mathtt{0\texttt{-}codom}} 

\renewcommand{\arraystretch}{1.3}  

\title{A parameterization process as a categorical construction} 
\author{
C\' esar Dom\'\i nguez \thanks{
Departamento de Matem\'aticas y Computaci\'on,
Universidad de La Rioja,
Edificio Vives, Luis de Ulloa s/n, E-26004 Logro\~no, La Rioja, Spain,
cesar.dominguez@unirioja.es.}
\and Dominique Duval \thanks{
Laboratoire Jean Kuntzmann, Universit\'e de Grenoble, 
51 rue des math\'ematiques, BP 53, F-38041 Grenoble C\'edex 9, France, 
Dominique.Duval@imag.fr.}
}
\date{August 25., 2009}

\begin{document}

\maketitle

\begin{itemize}
\item[] \textbf{Abstract.}
The parameterization process used in the symbolic computation systems 
Kenzo and EAT is studied here as a general construction in a categorical framework.
This parameterization process starts from a given specification 
and builds a parameterized specification 
by transforming some operations into parameterized operations,
which depend on one additional variable called the parameter. 
Given a model of the parameterized specification, 
each interpretation of the parameter, called an argument, 
provides a model of the given specification.
Moreover, under some relevant terminality assumption, 
this correspondence between the arguments and the models of the 
given specification is a bijection. 
It is proved in this paper that the parameterization process 
is provided by a free functor
and the subsequent parameter passing process by a natural transformation.
Various categorical notions are used, mainly  
adjoint functors, pushouts and lax colimits.
\end{itemize}

\section{Introduction}


Kenzo \cite{Kenzo} and its predecessor EAT \cite{EAT} are software
systems developed by F. Sergeraert. They are devoted to
Symbolic Computation in Algebraic Topology. In particular, they 
carry out calculations of homology groups of complex topological
spaces, namely iterated loop spaces. By means of EAT and Kenzo,
some homology groups that had never been obtained with any other
method, neither theoretical nor automatic, have been computed. In
view of the obtained results, some years ago, the first author 
of this paper began the formal study of the programs, in order to
reach a good understanding on the internal calculation processes
of these software systems.
In particular, our study of the data types used in EAT and
Kenzo \cite{LPR03,DLR07,DRS06} 
shows that there are two different layers of data structures in the systems.
In the first layer, one finds the usual abstract data types,
like the type of integers. 
In the second layer, one deals with algebraic structures,
like the structure of groups, 
which are implemented thanks to the abstract data types belonging to the first layer. 
In addition, we realized that in a system such as EAT, 
we do not simply implement one group, 
but more generally \emph{parameterized} families of groups. 
In \cite{LPR03} an operation is defined, 
which is called the \emph{imp} construction
because of its role in the implementation process in the system EAT.
Starting from a specification $\Ss$
in which some operations are labelled as ``pure'' \cite{DRS06}, 
the \emph{imp} construction builds a new specification $\Ss_A$ 
with a distinguished sort $A$
which is added to the domain of each non-pure operation.
It follows that each implementation of $\Ss_A$ 
defines a family of implementations of $\Ss$ 
depending on the choice of a value in the interpretation of $A$.
Besides, working with the \emph{imp} construction in \cite{LPR03} we were able
to prove that the implementations of EAT algebraic structures are as general as
possible, in the sense that they are ingredients of terminal objects in
certain categories of models;
this result is called the \emph{exact parameterization property}.
Later on, led by this characterization of EAT algebraic structures, 
in \cite{LPR03} we reinterpreted our results in terms of object-oriented technologies
like hidden algebras \cite{GM00} or coalgebras \cite{Ru00}.


This paper deals with generalization by parameterization in the sense of Kenzo and EAT,
so that our \emph{parameters} are symbolic constants of a given type, 
that will be replaced by \emph{arguments} which are elements in a given set.
The notion of parameterization in programming and specification languages 
bears several meanings, where the parameter may be a type or a specification.
For instance, in object-oriented programming, 
parametric polymorphism is called generic programming, 
in C++ it is characterized by the use of template parameters 
to represent abstract data types. 
On the other hand, in algebraic specifications, 
a parameterized specification is defined as a morphism of specifications
where the parameter is the source 
and the parameter passing is defined as a pushout \cite{ADJ80}. 


The framework for this paper is provided by \emph{equational logic}, 
considered from a categorical point of view.  
An equational theory, or simply a theory, is a category with chosen finite products. 
A model $M$ of a theory $\Tt$ is a functor $M\colon \Tt\to\Set$ which maps 
the chosen products to cartesian products. 
A theory $\Tt$ can be presented by a specification $\Ss$,
this means that $\Ss$ generates $\Tt$. 
In this paper, we are not interested in specifications for themselves, 
but as presentations of theories.
So, specifications are used mainly in the examples, 
and we feel free to modify a specification whenever needed 
as long as the presented theory is not changed.


The \emph{parameterization process} studied in this paper 
is essentially the ``\emph{imp} construction'' of \cite{LPR03}.
Starting from a theory $\Tt$ it provides a \emph{parameterized theory}~$\Tt_A$
by adding a \emph{type of parameters} $A$ 
and by transforming each term $f\colon X\to Y$ in $\Tt$ into 
a parameterized term $f'\colon A\times X\to Y$ in $\Tt_A$.
Then clearly $\Tt_A$ generalizes $\Tt$:
the models of $\Tt$ can be identified to the models of $\Tt_A$
which interpret the type of parameters $A$ as a singleton. 
There is another way to relate $\Tt$ and $\Tt_A$,
called the \emph{parameter passing process}, which runs as follows.
By adding to $\Tt_A$ a constant  $a$ (called the \emph{parameter}) of type $A$ we get 
a \emph{theory with parameter}~$\Tt_a$, 
such that for each parameterized term $f'\colon A\times X\to Y$ in $\Tt_A$
there is a term $f'(a,-)\colon X\to Y$ in $\Tt_a$.
Then the \emph{parameter passing morphism} $j\colon \Tt \to \Tt_a$ 
maps each term $f\colon X\to Y$ in $\Tt$ to $f'(a,-)\colon X\to Y$ in $\Tt_a$.
Given a model $M_A$ of $\Tt_A$ 
an \emph{argument} $\alpha$ is an element of the set $M_A(A)$,
it provides a model $M_{A,\alpha}$ of $\Tt_a$ 
which extends $M_A$ and satisfies $M_{A,\alpha}(a)=\alpha$.
Thanks to the parameter passing morphism, 
the model $M_{A,\alpha}$ of $\Tt_a$ gives rise to a model $M$ of $\Tt$ 
such that $M(f)=M_A(f')(\alpha,-)$ for each term $f$ in $\Tt$.
Moreover, under some relevant terminality assumption on $M_A$, 
this correspondence between the arguments $\alpha\in M_A(A)$ 
and the models of $\Tt$ is a bijection: 
this is the \emph{exact parameterization property}.


The parameterization process and its associated parameter passing process
have been described for each given theory $\Tt$,
but in fact they have the property of preserving the theory structure, 
which can be stated precisely in a categorical framework:
this is the aim of this paper. 
The parameterization process is defined as a \emph{functor}:
the construction of the parameterized theory~$\Tt_A$ from the given theory~$\Tt$ is 
a functor left adjoint to the construction of a coKleisli category,
and more precisely it is a free functor in the sense of section~\ref{sec:free}.
The parameter passing process is defined as a \emph{natural transformation}, 
along the following lines.  
First, the construction of the theory with parameter~$\Tt_a$
from the parameterized theory~$\Tt_A$ is simply a pushout construction,
such that the construction of~$\Tt_a$ from~$\Tt$ is a functor.
Then, each parameter passing morphism $j:\Tt\to\Tt_A$ is defined 
from a lax colimit of theories,
in such a way that the parameter passing morphisms are (essentially) the components 
of a natural transformation from the identity functor to this functor.


A first version of this approach can be found in \cite{DDLR05}, 
it relies on \emph{diagrammatic logic} \cite{Du03,Du07}.
In this paper, the explicit use of diagrammatic logic is postponed to the appendix.
With respect to the previous papers like \cite{LPR03}, 
we provide a new interpretation of the parameterization process 
and in addition an interpretation of the parameter passing process.
Moreover, we take into account the fact that there is a pure part in the given theory,
and we derive the exact parameterization property
from a more general result which does not rely on the existence of a terminal model.


In section~\ref{sec:defi} equational theories are defined 
and several examples are presented.
The parameterization process and the parameter passing process
are defined categorically in section~\ref{sec:const}. 
In section~\ref{sec:free} free functors are defined 
as left adjoint functors associated to morphisms of limit sketches,
and it is proved that the parameterization functor is free. 
The diagrammatic point of view on equational logic is presented in appendix~\ref{sec:dia}.
Most of the categorical notions used in this paper can be found in \cite{MacLane98}
or in \cite{BW99}.
We omit the size issues: for instance most colimits should be small. 
A \emph{graph} always means a directed multigraph,
and in order to distinguish between various kinds of structures 
with an underlying graph, 
we speak about the \emph{objects} and \emph{morphisms} of a category, 
the \emph{types} and \emph{terms} of a theory (or a specification) 
and the \emph{points} and \emph{arrows} of a limit sketch.

\section{Examples and definitions} 
\label{sec:defi}

\subsection{Equational theories and specifications} 
\label{subsec:equa}

In this paper, equational logic is seen from a categorical point of view, 
as for instance in \cite{Pitts}. 

\begin{defi}
\label{defi:equa-thry}
The category $\catT_{\eq}$ of \emph{equational theories} 
is made of the categories with chosen finite products
together with the functors which preserve the chosen finite products.
In addition, $\catT_{\eq}$ can be seen as a 2-category 
with the natural transformations as 2-cells. 
\end{defi}

Equational theories are called simply \emph{theories}.
For instance, the theory $\Set$ is made of the category of sets 
with the cartesian products as chosen products.

\begin{rema}
\label{rema:equa-thry}
The correspondence between 
equational theories in the universal algebra style (as in \cite{LEW96}) 
and equational theories in the categorical style (as defined here)  
can be found in \cite{Pitts}. 
Basically, the \emph{sorts} and products of sorts become objects, still called types,
the \emph{operations} and \emph{terms} become morphisms, still called terms 
(the \emph{variables} correspond to projections, as in example~\ref{exam:sgp})
and the \emph{equations} become equalities:
for instance a commutative square $g_1\circ f_1=g_2\circ f_2$ 
means that there is a term $h$ such that $g_1\circ f_1=h$ and $g_2\circ f_2=h$.
However a more subtle point of view on equations is presented in appendix~\ref{sec:dia}.
\end{rema}

\begin{defi}
\label{defi:equa-mod}
A \emph{(strict) model} $M$ of a theory $\Tt$ is 
a morphism of theories $M\colon \Tt\to\Set$
and a \emph{morphism $m\colon M\to M'$ of models} of $\Tt$ is a natural transformation. 
This forms the category $\Mod(\Tt)$ of models of $\Tt$.
\end{defi}

For every morphism of equational theories $\tt\colon \Tt_1\to\Tt$,
we denote by $\tt^\md\colon \Mod(\Tt)\to\Mod(\Tt_1)$ 
the functor which maps each model $M$ of $\Tt$ 
to the model $\tt^\md(M)=M\circ \tt$ of $\Tt_1$
and each morphism $m\colon M\to M'$ to $m\circ \tt$. 
In addition, for each model $M_1$ of $\Tt_1$, 
the category of \emph{models of $\Tt$ over $M_1$} is denoted $\Mod(\Tt)|_{M_1}$,
it is the subcategory of $\Mod(\Tt)$ 
made of the models $M$ such that $\tt^\md(M)=M_1$
and the morphisms $m$ such that $\tt^\md(m)=\id_{M_1}$.
Whenever $\tt$ is surjective on types, the category $\Mod(\Tt)|_{M_1}$ is discrete.

A theory $\Tt$ can be described by some presentation: 
a \emph{presentation} of an equational theory $\Tt$
is an equational specification $\Ss$ which generates $\Tt$;
this is denoted $\Tt\dashv\Ss$.
Two specifications are called \emph{equivalent} when they present the same theory.
An equational specification can be defined either 
in the universal algebra style as a signature (made of sorts and operations)
together with equational axioms, or equivalently, in a more categorical style,  
as a finite product sketch, see \cite{Lellahi89}, \cite{BW99},
and also section~\ref{subsec:sketch} and appendix~\ref{subsec:dia-equ}.
The correspondence between the universal algebra and the categorical  
points of view runs as in remark~\ref{rema:equa-thry}.

\begin{defi}
\label{defi:equa-spec}
The category $\catS_{\eq}$ of \emph{equational specifications} 
is the category of finite product sketches.
With (generalized) natural transformations as 2-cells, $\catS_{\eq}$ can be seen as a 2-category.
\end{defi}

Equational specifications are called simply \emph{specifications}. 
The category $\catT_{\eq}$ can be identified to a subcategory of $\catS_{\eq}$
(more precisely, to a reflective subcategory of $\catS_{\eq}$).
When $\Ss$ is a presentation of $\Tt$, a model of $\Tt$
is determined by its restriction to $\Ss$, which is called a \emph{model} of $\Ss$,
and in fact $\Mod(\Tt)$
can be identified to the category $\Mod(\Ss)$ of models of $\Ss$.

\begin{figure}[!t]
$$ \begin{array}{|l|l|c|}
\hline
 & \textrm{subscript} \; \ttE & \Ss_{\ttE} \\
\hline
\hline
  \textrm{type (or sort)} & \Type & X \\ 
\hline
  \textrm{term (or operation)} & \Term & 
  \xymatrix@C=3pc{X\ar[r]^{f} & Y} \\ 
\hline
  \textrm{selection of identity} & \Selid &  
  \xymatrix@C=3pc{X\ar[r]^{\id_X} & X} \\
\hline
  \textrm{composition} & \Comp &  
  \xymatrix@C=3pc{X\ar[r]^{f} \ar@/_3ex/[rr]_{g\circ f}^{=} & Y\ar[r]^{g} & Z}  \\ 
\hline
  \textrm{binary product} & \bProd &   
  \xymatrix@C=3pc@R=.7pc{ X & \\ 
    & X\stimes Y \ar[lu]_{p_X} \ar[ld]^{p_Y} \\ Y & \\ } \\ 
\hline
  \textrm{pairing (or binary tuple)} & \bTuple &  
  \xymatrix@C=3pc@R=1pc{ & X & \\ 
    Z \ar[ru]^{f} \ar[rd]_{g} \ar[rr]|{\,\tuple{f,g}\,} & 
    \ar@{}[u]|{=}\ar@{}[d]|{=} & 
    X\stimes Y \ar[lu]_{p_X} \ar[ld]^{p_Y} \\ & Y & \\ } \\ 
\hline
\end{array}$$
\caption{\label{fig:elem} Elementary specifications}
\end{figure}

We will repeatedly use the fact that $\catT_{\eq}$ and $\catS_{\eq}$,
as well as other categories of theories and of specifications, have colimits,
and that left adjoint functors preserve colimits. 
In addition every specification is the colimit of a diagram of elementary specifications.
The \emph{elementary specifications} are the specifications respectively made of: 
a type, 
a term, 
an identity term, 
a composed term, 
a $n$-ary product 
and a $n$-ary tuple, for all $n\geq0$,
as in figure~\ref{fig:elem} (where only $n=2$ is represented).
Let us consider a theory $\Tt$ presented by a specification $\Ss$, 
then $\Ss$ is the colimit of a diagram $\Delta$ of elementary specifications,
and $\Tt$ is the colimit of the diagram of theories generated by $\Delta$.

\subsection{Examples} 
\label{subsec:exam}

\begin{exam}
\label{exam:term}

Let us consider the theory $\Tt_{\oper,0}$ presented by two types $X,Y$,
and the three following theories extending $\Tt_{\oper,0}$ 
(the subscript $\oper$ stands for ``operation'',
since $\Tt_{\oper}$ is presented by the elementary specification 
for terms or operations $\Ss_{\Term}$). 
The unit type is denoted $\uno$ and the projections are not given any name.

$$  \begin{array}{c|c|c|c|}
\cline{2-2} 
  \Tt_{\oper,A} \dashv  & 
  \xymatrix@R=1pc{A & A\stimes X\ar[l]\ar[d]\ar[rd]^{f'} & \\ & X & Y \\ } &  
  \multicolumn{2}{c}{} \\
\cline{2-2} \cline{4-4} 
  \multicolumn{2}{c}{} & 
  \qquad \Tt_{\oper,a} \dashv  & 
  \xymatrix@R=1pc{A & A\stimes X\ar[l]\ar[d]\ar[rd]^{f'} & \\ \uno \ar[u]^{a} &X & Y \\ } \\
\cline{2-2} \cline{4-4} 
  \Tt_{\oper} \dashv  & 
  \xymatrix{ & X\ar[r]^{f} &Y \\ } & 
  \multicolumn{2}{c}{} \\
\cline{2-2} 
\end{array}$$

These theories are related by various morphisms (all of them preserving $\Tt_{\oper,0}$):
$\tt_{\oper,A}\colon \Tt_{\oper,A}\to\Tt_{\oper}$ maps $A$ to $\uno$
and $\tt_{\oper,a}\colon \Tt_{\oper,a}\to\Tt_{\oper}$ extends $\tt_{\oper,A}$ 
by mapping $a$ to $\id_{\uno}$,
while $j_{\oper,A}\colon \Tt_{\oper,A}\to\Tt_{\oper,a}$ is the inclusion.
In addition, here are two other presentations of the theory $\Tt_{\oper,a}$
(the projections are omitted and $ \uno\times X$ is identified to $X$):

$$ 
 \begin{array}{|c|}
\cline{1-1}
\xymatrix@R=1pc{A & A\stimes X\ar[rd]^{f'} & \\ 
\uno \ar[u]^{a} & X  \ar@{}[ru]|(.3){=} \ar[u]^{a\stimes\id_X} \ar[r]_{f''} & Y \\ } \\ 
\cline{1-1}
\end{array}  
\qquad\qquad 
\begin{array}{|c|}
\cline{1-1}
\xymatrix@R=1pc{A & A\stimes X\ar[r]^{f'} \ar@{}[rd]|{=} &Y \\ 
\uno \ar[u]^{a} & X \ar[u]^{a\stimes\id_X} \ar[r]_{f''} & Y \ar[u]_{\id_Y}  \\ } \\ 
\cline{1-1}
\end{array} $$
It is clear from anyone of these new presentations of $\Tt_{\oper,a}$
that there is a morphism $j_{\oper}\colon \Tt_{\oper}\to\Tt_{\oper,a}$ 
which maps $f$ to $f''$.
In addition, $\tt_{\oper,a}\circ j_{\oper,A} = \tt_{\oper,A}$
and there is a natural transformation
$t_{\oper} \colon  j_{\oper} \circ \tt_{\oper,A}  \To j_{\oper,A}$ 
defined by $(t_{\oper})_X=\id_X$, $(t_{\oper})_Y=\id_Y$ 
and $(t_{\oper})_A=a\colon \uno\to A$.

$$ \xymatrix@R=.8pc{
  \Tt_{\oper,A} \ar[dd]_{\tt_{\oper,A}} \\ 
  \mbox{ } \\ 
  \Tt_{\oper}  \\
  }
\qquad \qquad 
\xymatrix@R=.8pc{
  \Tt_{\oper,A} \ar[dr]^{j_{\oper,A}} \ar[dd]_{\tt_{\oper,A}} \\ 
  \ar@{}[r]|(.4){=} & \Tt_{\oper,a} \ar[dl]^{\tt_{\oper,a}} \\ 
  \Tt_{\oper}  & \\
  }
\qquad \qquad 
 \xymatrix@R=.8pc{
  \Tt_{\oper,A} \ar[dr]^{j_{\oper,A}} \ar[dd]_{\tt_{\oper,A}} & \\ 
  \ar@{}[r]|(.4){\trnat}|(.25){t_{\oper}} & \Tt_{\oper,a}  \\ 
  \Tt_{\oper} \ar[ur]_{j_{\oper}} \\ 
  }
$$ 

\textbf{Parameterization process}
(construction of $\Tt_{\oper,A}$ from $\Tt_{\oper}$).
The theory $\Tt_{\oper,A}$ is obtained from $\Tt_{\oper}$
by adding a type $A$, called the \emph{type of parameters},
to the domain of the unique term in $\Tt_{\oper}$.
Then $\Tt_{\oper,A}$ can be seen as a \emph{generalization} of $\Tt_{\oper}$,
since each model $M$ of $\Tt_{\oper}$ can be identified to a model of $\Tt_{\oper,A}$
where $M(A)$ is a singleton. 
We will also say that $\Tt_{\oper,A}$ is the \emph{expansion} of $\Tt_{\oper}$.

\textbf{Parameter passing process}  
(construction of $\Tt_{\oper,a}$ from $\Tt_{\oper,A}$
and of a morphism from $\Tt_{\oper}$ to $\Tt_{\oper,a}$). 
The theory $\Tt_{\oper,a}$ is obtained from $\Tt_{\oper,A}$
by adding a constant term $a\colon \uno\to A$, called the \emph{parameter}.
A model $M_a$ of $\Tt_{\oper,a}$ is made of a model $M_A$ of $\Tt_{\oper,A}$
together with an element $\alpha=M_a(a)\in M_A(A)$,
so that we can denote $M_a=(M_A,\alpha)$.
Now, let $M_A$ be some fixed model of $\Tt_{\oper,A}$,
then the models $M_a$ of $\Tt_{\oper,a}$ over $M_A$ correspond bijectively 
to the elements of $M_A(A)$ by $M_a \mapsto M_a(a)$,
so that we get the \emph{parameter adding} bijection
(the category $\Mod(\Tt_{\oper,a})|_{M_A}$ is discrete):
  $$ \Mod(\Tt_{\oper,a})|_{M_A} \isoto M_A(A) \hsp \mbox{by} \hsp  
  M_a = (M_A,\alpha) \mapsto M_a(a)=\alpha \;. $$
On the other hand, 
each model $M_a=(M_A,\alpha)$ of $\Tt_{\oper,a}$ gives rise to a model
${j_{\oper}}^\md(M_a)$ of $\Tt_{\oper}$ such that 
${j_{\oper}}^\md(M_a)(X)=M_a(X)=M_A(X)$, ${j_{\oper}}^\md(M_a)(Y)=M_a(Y)=M_A(Y)$
and ${j_{\oper}}^\md(M_a)(f)=M_a(f'')=M_A(f')(\alpha,-) $.
Now, let $M_A$ be some fixed model of $\Tt_{\oper,A}$ 
and $M_0$ its restriction to $\Tt_{\oper,0}$,
then for each model $M_a=(M_A,\alpha)$ of $\Tt_{\oper,a}$ over $M_A$ 
the model ${j_{\oper}}^\md(M_a)$ of $\Tt_{\oper}$ is over $M_0$.
This yields the \emph{parameter passing} function 
(the categories $\Mod(\Tt_{\oper,a})|_{M_A}$ and $\Mod(\Tt_{\oper})|_{M_0}$ 
are discrete):
  $$ \Mod(\Tt_{\oper,a})|_{M_A} \to \Mod(\Tt_{\oper})|_{M_0}  \hsp \mbox{by} \hsp  
  M_a  \mapsto {j_{\oper}}^\md(M_a) \;. $$

\textbf{Exact parameterization.} 
Let $M_0$ be any fixed model of $\Tt_{\oper,0}$,
it is made of two sets $\setX=M_0(X)$ and $\setY=M_0(Y)$. 
Let $M_A$ be the model of $\Tt_{\oper,A}$ over $M_0$ such that 
$M_A(A)=\setY^{\setX}$ and 
$M_A(f')\colon \setY^{\setX} \times \setX \to \setY$ 
is the application. 
It can be noted that $M_A$ is the terminal model of $\Tt_{\oper,A}$ over $M_0$. 
Then the parameter passing function is a bijection, 
and composing it with the parameter adding bijection we get
(where $\curry{M(f)}\in\setY^{\setX}$ corresponds by currying to 
$M(f)\colon \setX\to\setY$):
  $$ \Mod(\Tt_{\oper})|_{M_0} \iso M_A(A) \hsp \mbox{by} \hsp  
  M_{A,\alpha} \leftrightarrow \alpha  \hsp \mbox{i.e., by} \hsp 
  M \leftrightarrow \curry{M(f)} \;. $$ 

\end{exam}

\begin{exam}
\label{exam:sgp}
Let $\Tt_{\sgp}$ be the theory for semigroups 
presented by one type $G$, 
one term $\prd\colon G^2 \to G$ 
and one equation $\prd(x,\prd(y,z))=\prd(\prd(x,y),z)$
where $x$, $y$, $z$ are variables of type $G$.
As usual with the categorical point of view, 
in fact the \emph{variables} are projections; 
here, $x,y,z\colon G^3\to G$ are the three projections 
and for instance $\prd(x,y)$ is $\prd\circ\tuple{x,y}\colon G^3\to G$,
composed of the pair $\tuple{x,y}\colon G^3\to G^2$ and of $\prd\colon G^2 \to G$
(more details are given in the appendix, example~\ref{exam:dia-sg}).

\textbf{Parameterization process}. 
In order to get parameterized families of semigroups, 
we consider the theory $\Tt_{\sgp,A}$ 
presented by two types $A$ and $G$, 
one term $\prd'\colon A\times G^2 \to G$ 
and one equation $\prd'(p,x,\prd'(p,y,z))= \prd'(p,\prd'(p,x,y),z)$ 
where $x$, $y$, $z$ are variables of sort $G$ 
and $p$ is a variable of sort $A$. 

\textbf{Parameter passing process}.
The theory $\Tt_{\sgp,a}$ is $\Tt_{\sgp,A}$ together with a parameter $a\colon \uno\to A$,
hence with $\prd''=\prd'\circ(a\times\id_{G^2}) \colon G^2 \to G$ 
(where $\uno \times G^2$ is identified to $G^2$).
Each model $M_A$ of $\Tt_{\sgp,A}$ gives rise to a family of models of $\Tt_{\sgp,a}$,
all of them with the same underlying set $M_A(G)$
but with different interpretations of $a$ in $M_A(A)$.
Mapping $\prd$ to $\prd''$ defines a morphism from $\Tt_{\sgp}$ to $\Tt_{\sgp,a}$.
So, each model $M_a$ of $\Tt_{\sgp,a}$ gives rise to a model $M$ of $\Tt_{\sgp}$ such that 
$M(G)=M_a(G)$ and $M(\prd)(x,y)=M_a(\prd')(\alpha,x,y) $ for each $x,y\in M_a(G)$,
where $\alpha=M_a(a)$ is the \emph{argument}.

\end{exam}

\begin{exam}
\label{exam:list}
This example motivates the existence of pure terms in the given theory.
Let us consider the theory $\Tt_{\nat}$ ``of naturals''
presented by a type $N$ and two terms $z\colon \uno\to N$ and $s\colon N\to N$,
and let us say that $z$ is pure. 
Let $\Tt_{\nat,0}$ be the subtheory presented by $N$ and $z$,
it is called the pure subtheory of $\Tt_{\nat}$. 
We define the theory $\Tt_{\nat,A}$ as made of 
two types $A$ and $N$ and two terms 
$z\colon \uno\to N$ and $s'\colon A\times N\to N$.
It should be noted that $\Tt_{\nat,A}$ contains $\eps_\uno\colon A\times \uno\to \uno$ 
and $z'=z\circ \eps_\uno\colon A\times \uno\to N$.
Then $\Tt_{\nat,A}$ is a theory ``of lists of $A$'', with $z$ for the empty list 
and $s'$ for concatenating an element to a list.
In this way, the theory of lists of $A$ is built as a generalization 
of the theory of naturals; 
indeed the naturals can be identified to the lists over a singleton. 
\end{exam}

\begin{exam}
\label{exam:dm}
Here is another example where pure terms are required,
this is a simplified version of many structures in Kenzo/EAT.
Let $\Tt_{\mon}$ be the theory for monoids presented by
one type $G$,
two terms $\prd\colon G^2 \to G$ and $\unt\colon \to G$,
and the equations $\prd(x,\prd(y,z)) = \prd(\prd(x,y),z)$,
$\prd(x,\unt) = x$, $\prd(\unt,x) = x$ where $x$, $y$, $z$ are variables of type $G$.
Let $\Tt_{\dm}$ be the theory for \emph{differential monoids},
presented by $\Tt_{\mon}$ together with one term $\dif \colon G \to G$ 
and the equations $\dif(\prd(x,y)) = \prd(\dif(x),\dif(y))$, $\dif(\unt) = \unt$, $\dif(\dif(x)) = \unt$,
and with the terms in $\Tt_{\mon}$ as its pure terms. 
In order to get parameterized families of differential structures on one monoid,
we define the theory $\Tt_{\dm,A}$ 
presented by two types $G$, $A$ and three terms
$\prd\colon G^2 \to G$, $\unt\colon \uno\to G$ and $\dif' \colon A\times G \to G$
and the equations $\prd(x,\prd(y,z)) = \prd(\prd(x,y),z)$,
$\prd(x,\unt) = x$, $\prd(\unt,x) = x$, 
$\dif'(p,(\prd(x,y))) = \prd(\dif'(p,x), \dif'(p,y))$, $\dif'(p, \unt) = \unt$,
$\dif'(p,\dif'(p,x)) = \unt$.
Each model $M_A$ of $\Tt_{\dm,A}$ gives rise to a family
of models of $\Tt_{\dm}$, 
all of them with the same underlying monoid $(M_A(G),M_A(\prd),M_A(\unt))$:
there is a model $M_a$ of $\Tt_{\dm}$ over $M_A$ for each element $\alpha$ in $M_A(A)$,
with its differential structure defined by
$M_a(\dif)= M_A(\dif')(\alpha,-)$.
\end{exam}

\begin{exam}
\label{exam:pi}
In the next sections we will use the theories with the following presentations:
$$  \begin{array}{c|c|c|c|}
\cline{2-2} 
\Pi_A \dashv  & \xymatrix{A} &  \multicolumn{2}{c}{} \\
\cline{2-2} \cline{4-4}
\multicolumn{2}{c}{} & \qquad \Pi_a \dashv  & \xymatrix@R=1pc{ A \\ \uno \ar[u]^{a} \\ } \\
\cline{2-2} \cline{4-4}
\Pi \dashv  & \xymatrix{\uno} &  \multicolumn{2}{c}{} \\
\cline{2-2} 
\end{array}$$
These theories are related by several morphisms:
$\pi_A\colon  \Pi_A\to \Pi$ maps $A$ to $\uno$, 
both $i\colon \Pi\to \Pi_a$ and $i_A\colon  \Pi_A\to \Pi_a$ are the inclusions,
and $\pi_a\colon  \Pi_a\to \Pi$ extends $\pi_A$ by mapping $a$ to $\id_{\uno}$,
so that $\pi_A$ and $\pi_a$ are epimorphisms.
In addition, $\pi_a\circ i_A=\pi_A$ and there is a natural transformation
$p\colon  i \circ \pi_A  \To i_A$ defined by $p_A=a\colon \uno\to A$. 
The diagram below on the right is the \emph{lax colimit of $\pi_A$}, 
which means that it enjoys the following universal property:
for each $\Pi'_a$ with $i'_A\colon  \Pi_A\to \Pi'_a$, $i'\colon  \Pi\to \Pi'_a$ 
and $p'\colon i' \circ \pi_A  \To i'_A$,
there is a unique $h\colon  \Pi_a\to \Pi'_a$ such that 
$h\circ i_A = i'_A$, $h\circ i = i'$ and $h\circ p=p'$. 
For instance, given $\Pi$, $\pi_A\colon  \Pi_A\to \Pi$, $\id_{\Pi}\colon  \Pi\to \Pi$ 
and $\id_{\pi_A}\colon \pi_A \To \pi_A$,
then $\pi_a\colon  \Pi_a\to \Pi$ is the unique morphism such that 
$\pi_a\circ i_A = \pi_A$, $\pi_a\circ i = \id_{\Pi}$ and $\pi_a\circ p=\id_{\pi_A}$.
$$ \xymatrix@R=.8pc{
  \Pi_A \ar[dd]_{\pi_A} \\ 
  \\ 
  \Pi  \\
  }
\qquad \qquad 
\xymatrix@R=.8pc{
  \Pi_A \ar[dr]^{i_A} \ar[dd]_{\pi_A} \\ 
  \ar@{}[r]|(.4){=} & \Pi_a \ar[dl]^{\pi_a} \\ 
  \Pi  & \\
  }
\qquad \qquad 
 \xymatrix@R=.8pc{
  \Pi_A \ar[dr]^{i_A} \ar[dd]_{\pi_A} & \\ 
  \ar@{}[r]|(.4){\trnat}|(.25){p} & \Pi_a  \\ 
  \Pi \ar[ur]_{i} \\ 
  }
$$ 
\end{exam}

\subsection{Some other kinds of theories} 
\label{subsec:thry}

For every theory $\Tt$, 
the coslice category of \emph{theories under $\Tt$} is denoted $\Tt\cosl\catT_{\eq}$.
It can be seen as a 2-category, with 
the natural transformations which extend the identity on $\Tt$ as 2-cells.

\begin{defi}
\label{defi:A-th}
A \emph{parameterized theory} $\Tt_A$ 
is a theory $\Tt$ with a distinguished type, 
called the \emph{type of parameters} and usually denoted $A$. 
The 2-category of parameterized theories 
is the coslice 2-category $\catT_A=\Pi_A\cosl\catT_{\eq}$ of theories under $\Pi_A$.
A \emph{theory with a parameter} $\Tt_a$ 
is a parameterized theory with a distinguished constant of type $A$,
called the \emph{parameter} and usually denoted $a\colon \uno\to A$.
The 2-category of theories with a parameter 
is the coslice 2-category $\catT_a=\Pi_a\cosl\catT_{\eq}$ of theories under $\Pi_a$.
\end{defi}

According to the context,
$\Tt_A$ usually denotes the parameterized theory $\gamma_A \colon \Pi_A\to\Tt_A$,
and sometimes it denotes the equational theory $\Tt_A$ itself. 
Similarly for $\Tt_a$, which usually denotes $\gamma_a\colon \Pi_a\to\Tt_a$
and sometimes $\Tt_a$ itself. 

In addition, it can be noted that $\Pi$ is the initial theory
(which may also be presented by the empty specification) 
so that $\Pi\cosl\catT_{\eq}$ is isomorphic to $\catT_{\eq}$.

The 2-categories $\catS_A$ and $\catS_a$ of \emph{parameterized specifications}
and \emph{specifications with a parameter}, respectively, 
are defined in a similar way.

On the other hand, the input of the parameterization process 
is a theory $\Tt$ together with a wide subtheory $\Tt_0$
(\emph{wide} means: with the same types), 
such a structure is called a decorated theory.

\begin{defi}
\label{defi:dec-th}
A \emph{decorated theory} is made of a theory $\Tt$ with a wide subtheory $\Tt_0$ 
called the \emph{pure} subtheory of $\Tt$.
A morphism of decorated theories is a morphism of theories $\tt\colon \Tt\to\Tt'$ 
which maps the pure part of $\Tt$ to the pure part of $\Tt'$.
This forms the category $\catT_{\dec}$ of decorated theories.  
\end{defi}

So, a decorated theory $\Tt$ is endowed with a distinguished family of terms,
called the \emph{pure} terms, 
such that all the identities and projections are pure  
and every composition or tuple of pure terms is pure.
Pure terms are denoted with ``$\rpto$''.
When there is no ambiguity we often use the same notation $\Tt$ 
for the theory $\Tt$ itself  
and for the decorated theory made of $\Tt$ and $\Tt_0$.

The decorated specifications are defined in a straightforward way.
For instance, we may consider the decorated specification made of 
a type $N$, a pure term $z\colon \uno\rpto N$ and a term $s\colon N\to N$
(see example~\ref{exam:list}).

\section{Constructions} 
\label{sec:const}

\subsection{The parameterization process is a functor} 
\label{subsec:gene}

In this section we prove that the parameterization process is functorial,
by defining a functor $F_{\expan}\colon \catT_{\dec}\to\catT_A$
(``$\expan$'' for ``expansion'') 
which adds the type of parameters to the domain of every non-pure term.
In addition, theorem~\ref{theo:gene} states that $F_{\expan}$ is left adjoint to 
the functor $G_{\expan}\colon \catT_A\to\catT_{\dec}$, 
which builds the coKleisli category of the comonad $A\times-$.
Moreover, we will see in section~\ref{sec:free} that $F_{\expan}$ is a free functor 
associated to a morphism of limit sketches, 
and in appendix~\ref{sec:dia} that this morphism of limit sketches 
underlies a morphism of diagrammatic logics. 
 

\begin{figure}[!t]
$$ \begin{array}{|l|l|c|c|}
\hline
 & \textrm{index} \; \ttE\deco & \Ss_{\ttE\deco} & F_{\expan}\Ss_{\ttE\deco} \\
\hline
\hline
  \textrm{type} & \Type\pur & X & X \\ 
\hline
  \textrm{pure term} & \Term\pur & 
  \xymatrix@C=3pc{X\ar@{~>}[r]^{f} & Y} & 
  \xymatrix@C=3pc{X\ar[r]^{f} & Y} \\ 
\hline
  \textrm{term} & \Term\gen &
  \xymatrix@C=3pc{X\ar[r]^{f} & Y} & 
  \xymatrix@C=3pc{A\stimes X\ar[r]^{f'} & Y}  \\ 
\hline
  \textrm{sel. of identity} & \Selid\pur &
  \xymatrix@C=3pc{X\ar@{~>}[r]^{\id_X} & X} & 
  \xymatrix@C=3pc{X\ar[r]^{\id_X} & X} \\
\hline
  \textrm{pure composition} & \Comp\pur & 
  \xymatrix@C=3pc{X\ar@{~>}[r]^{f} \ar@/_3ex/@{~>}[rr]_{g\circ f}^{=} & Y\ar@{~>}[r]^{g} & Z} & 
  \xymatrix@C=3pc{X\ar[r]^{f} \ar@/_3ex/[rr]_{g\circ f}^{=} & Y\ar[r]^{g} & Z} \\
\hline
  \textrm{composition} & \Comp\gen &  
  \xymatrix@C=3pc{X\ar[r]^{f} \ar@/_3ex/[rr]_{g\circ f}^{=} & Y\ar[r]^{g} & Z} & 
  \xymatrix@C=3pc{A\stimes X \ar[r]^{\tuple{\proj_X,f'}} 
     \ar@/_3ex/[rr]_{g'\circ\tuple{\proj_X,f'}}^{=} & A\stimes Y\ar[r]^{g'} & Z \\} \\ 
\hline
  \textrm{binary product} & \bProd\pur &  
  \xymatrix@C=3pc@R=.7pc{ X & \\ 
    & X\stimes Y \ar@{~>}[lu]_{p_X} \ar@{~>}[ld]^{p_Y} \\ Y & \\ }  & 
  \xymatrix@C=3pc@R=.7pc{ X & \\ 
    & X\stimes Y \ar[lu]_{p_X} \ar[ld]^{p_Y} \\ Y & \\ } \\ 
\hline
  \textrm{pure pairing} & \bTuple\pur & 
  \xymatrix@C=3pc@R=1pc{ & X & \\ 
    Z \ar@{~>}[ru]^{f} \ar@{~>}[rd]_{g} \ar@{~>}[rr]|{\tuple{f,g}} & \ar@{}[u]|{=}\ar@{}[d]|{=} & 
    X\stimes Y \ar@{~>}[lu]_{p_X} \ar@{~>}[ld]^{p_Y} \\ & Y & \\ }  & 
  \xymatrix@C=3pc@R=1pc{ & X & \\ 
    Z \ar[ru]^{f} \ar[rd]_{g} \ar[rr]|{\tuple{f,g}} & 
    \ar@{}[u]|{=}\ar@{}[d]|{=} & 
    X\stimes Y \ar[lu]_{p_X} \ar[ld]^{p_Y} \\ & Y & \\ } \\ 
\hline
  \textrm{pairing} & \bTuple\gen & 
  \xymatrix@C=3pc@R=1pc{ & X & \\ 
    Z \ar[ru]^{f} \ar[rd]_{g} \ar[rr]|{\tuple{f,g}} & \ar@{}[u]|{=}\ar@{}[d]|{=} & 
    X\stimes Y \ar@{~>}[lu]_{p_X} \ar@{~>}[ld]^{p_Y} \\ & Y & \\ }  & 
  \xymatrix@C=3pc@R=1pc{ & X & \\ 
    A\stimes Z \ar[ru]^{f'} \ar[rd]_{g'} \ar[rr]|{\tuple{f',g'}} & 
    \ar@{}[u]|{=}\ar@{}[d]|{=} & 
    X\stimes Y \ar[lu]_{p_X} \ar[ld]^{p_Y} \\ & Y & \\ } \\ 
\hline
\end{array}$$
\caption{\label{fig:F-elem} The functor $F_{\expan}$ 
on elementary decorated specifications}
\end{figure}

In order to define the functor $F_{\expan}$ 
we use the fact that it should preserve colimits.
It has been seen in section~\ref{subsec:equa} that 
every specification is the colimit of a diagram of elementary specifications.
Similarly, every decorated specification is the colimit of 
a diagram of elementary decorated specifications, 
denoted $\Ss_{\ttE\deco}$ where 
$\mathtt{x}=\mathtt{p}$ for ``pure'' or $\mathtt{x}=\mathtt{g}$ for ``general''. 
Informally, the functor $F_{\expan}$ explicits the fact that every 
general feature  in a decorated specification gets parameterized, 
while every pure feature remains unparameterized.
Figure~\ref{fig:F-elem} defines the parameterized specification 
$F_{\expan}\Ss_{\ttE\deco}$ 
for each elementary decorated specification $\Ss_{\ttE\deco}$
(many projection arrows are omitted, and 
when needed the projections from $A\times X$ are denoted 
$\proj_X\colon A\times X\to A$ and $\eps_X\colon A\times X\to X$).
The morphisms of parameterized specifications $F_{\expan}\ss$,
for $\ss$ between elementary decorated specifications, 
are straightforward. 
For instance, let $c\colon \Ss_{\Term\gen}\to\Ss_{\Term\pur}$
be the conversion morphism,
which corresponds to the fact that every pure term can be seen as a general term,
then $F_{\expan}c$ maps $f'\colon A\times X\to Y$ in $F_{\expan}\Ss_{\Term\gen}$ 
to $f\circ \eps_X\colon A\times X\to Y$ in $F_{\expan}\Ss_{\Term\pur}$.
Now, given a decorated theory $\Tt$ presented by 
the colimit of a diagram $\Delta$ of elementary decorated specifications,
we define $F_{\expan}\Tt$ as the parameterized theory presented by 
the colimit of the diagram $F_{\expan}\Delta$ of parameterized specifications.

\begin{defi}
\label{defi:gene}
The functor $F_{\expan}:\catT_{\dec}\to\catT_A$ defined above 
is called the \emph{parameterization functor}.
\end{defi}

Clearly the parameterization functor preserves colimits. 
In addition, let $\Tt_A$ be the parameterized theory $F_{\expan}\Tt$,
it follows from the definition of $F_{\expan}$ that the equational theory $\Tt_A$
is a theory under $\Tt_0$. 


Now the functor $G_{\expan}$ is defined independently from $F_{\expan}$.
Let $\Tt_A$ be a parameterized theory.
The endofunctor of product with $A$ forms a comonad on $\Tt_A$
with the counit $\eps$ made of the projections $\eps_X\colon A\times X\to X$
and the comultiplication made of the terms 
$\delta_X\colon A\times X\to A\times A\times X$ induced by the diagonal on $A$.
Let $\Tt$ be the coKleisli category of this comonad:
it has the same types as $\Tt_A$
and a term $\kl{f}\colon X\to Y$ for each term $f\colon A\times X\to Y$ in $\Tt_A$. 
There is a functor from $\Tt_A$ to $\Tt$ which is the identity on types 
and maps every $g\colon X\to Y$ in $\Tt_A$ 
to $\kl{g\circ \eps_X}\colon X\to Y$ in $\Tt$. 
Then every finite product in $\Tt_A$ is mapped to a finite product in $\Tt$,
which makes $\Tt$ a theory. 
Let $\Tt_0$ denote the image of $\Tt_A$ in $\Tt$, it is a wide subtheory of $\Tt$.
In this way, any parameterized theory yields a decorated theory.
The definition of $G_{\expan}$ on morphisms is straightforward. 
The next result can be derived directly, 
or as a consequence of theorem~\ref{theo:free}.

\begin{theo}
\label{theo:gene}
The parameterization functor $F_{\expan}$ and the functor $G_{\expan}$ 
form an adjunction $F_{\expan}\dashv G_{\expan}$:
  $$ \xymatrix@C=4pc{ \catT_{\dec} \ar@/^/[r]^{F_{\expan}} \ar@{}[r]|{\bot} & 
  \catT_A \ar@/^/[l]^{G_{\expan}} } $$
\end{theo}

The next result states that $\Tt$ can be easily recovered from $\Tt_A$, 
by mapping $A$ to $\uno$.

\begin{prop}
\label{prop:general}
Let $\Tt$ be a decorated theory with pure subtheory $\Tt_0$
and $\gamma_A\colon \Pi_A\to\Tt_A$ the parameterized theory $F_{\expan}\Tt$.
Let $\gamma\colon \Pi\to\Tt$ be the unique morphism from the initial theory $\Pi$ 
to the theory $\Tt$.
Then there is a morphism $\tt_A\colon \Tt_A\to\Tt$ under $\Tt_0$
such that the following square is a pushout:
$$ \xymatrix{ 
  \ar@{}[rd]|{\PO} 
  \Pi_A \ar[d]_{\pi_A} \ar[r]^{\gamma_A} & \Tt_A \ar[d]^{\tt_A} \\ 
  \Pi \ar[r]^{\gamma} &\Tt \\  
  } $$
\end{prop}

\begin{proof}
It can easily be checked that this property is satisfied by each elementary specification.
Then the result follows by commuting two colimits:
on the one hand the colimit that defines the given theory from its elementary components,
and on the other hand the pushout.
\end{proof}

When there is an epimorphism of theories $\tt\colon \Tt_1\to\Tt_2$,
we say that $\Tt_1$ is \emph{the generalization of $\Tt_2$ along $\tt$}.
Indeed, since $\tt$ is an epimorphism, 
the functor $\tt^\md\colon \Mod(\Tt_2)\to\Mod(\Tt_1)$ is a monomorphism,
which can be used for identifying $\Mod(\Tt_2)$ to a subcategory of $\Mod(\Tt_1)$.

\begin{coro}
\label{coro:general}
With notations as in proposition~\ref{prop:general},
$\Tt_A$ is the generalization of $\Tt$ along $\tt_A$.
\end{coro}

\begin{proof}
Clearly $\pi_A\colon \Pi_A\to\Pi$ is an epimorphism.
Since epimorphisms are stable under pushouts,
proposition~\ref{prop:general} proves that $\tt_A\colon \Tt_A\to\Tt$ 
is also an epimorphism.
\end{proof}

Let $F_{\expan}:\catT_{\dec}\to\catT_A$ be the parameterization functor
and let $U\colon \catT_A\to\catT_{\eq}$ be the functor 
which simply forgets that the type $A$ is distinguished,
so that $U \circ F_{\expan}\colon \catT_{\dec}\to\catT_{\eq}$ 
maps the decorated theory $\Tt$ to the equational theory $\Tt_A$.
 $$ \xymatrix@C=4pc{ \catT_{\dec} \ar[r]^{F_{\expan}} & 
  \catT_A  \ar[r]^{U} &  \catT_{\eq} \\ } $$
Every theory $\Tt$ can be seen as a decorated theory where 
the pure terms are defined inductively as the identities, the projections, 
and the compositions and tuples of pure terms.
Let $I\colon \catT_{\eq}\to\catT_{\dec}$ denote the corresponding inclusion functor.
Then the endofunctor $U \circ F_{\expan}\circ I\colon \catT_{\eq}\to\catT_{\eq}$ 
corresponds to the ``\emph{imp} construction'' of \cite{LPR03},
which transforms each term $f\colon X\to Y$ in $\Tt$ into 
$f'\colon A\times X\to Y$ for a new type $A$.

\subsection{The parameter passing process is a natural transformation}
\label{subsec:passing}


A theory $\Tt_a$ with a parameter is built simply by adding 
a constant $a$ of type $A$ to a parameterized theory $\Tt_A$. 
Obviously, this can be seen as a pushout. 

\begin{defi}
\label{defi:adding}
Let $\gamma_A\colon \Pi_A\to\Tt_A$ be a parameterized theory.
The theory with parameter \emph{extending} $\gamma_A$ 
is $\gamma_a\colon \Pi_a\to\Tt_a$ given by the pushout of $\gamma_A$ and $i_A$:
$$ \xymatrix{ 
  \ar@{}[rd]|{\PO} 
  \Pi_A \ar[d]_{i_A} \ar[r]^{\gamma_A} & \Tt_A \ar[d]^{j_A} \\ 
  \Pi_a \ar[r]^{\gamma_a} &\Tt_a \\  
  } $$
\end{defi}

The pushout of theories in definition~\ref{defi:adding} 
gives rise to a pullback of categories of models, 
hence for each model $M_A$ of $\Tt_A$ 
the function which maps each model $M_a$ of $\Tt_a$ over $M_A$
to the element $M_a(a)\in M_A(A)$ defines a bijection:
\begin{equation}
\label{eq:adding}
 \Mod(\Tt_a)|_{M_A}  \isoto M_A(A) \;. 
\end{equation}

Let us assume that the parameterized theory $\gamma_A\colon \Pi_A\to\Tt_A$ 
is $F_{\expan}\Tt$ for some decorated theory $\Tt$
with pure subtheory $\Tt_0$.
Then the pushout property in definition~\ref{defi:adding} 
ensures the existence of a unique $\tt_a\colon  \Tt_a\to \Tt$
such that $\tt_a \circ \gamma_a = \gamma\circ\pi_a$
(which means that $\tt_a$ maps $A$ to $\uno$ and $a$ to $\id_{\uno}$)
and $\tt_a \circ j_A = \tt_A$.
Then $\Tt_A$ is a theory under $\Tt_0$ 
and the composition by $j_A$ makes $\Tt_a$ a theory under $\Tt_0$ 
with $j_A$ preserving~$\Tt_0$.

$$ \xymatrix@R=.8pc{
  \Tt_A \ar[dr]^{j_A} \ar[dd]_{\tt_A} & \\
  \ar@{}[r]|(.4){=} & \Tt_a \ar[dl]^{\tt_a} \\ 
  \Tt & \\ 
} $$


Lax cocones and lax colimits in 2-categories generalize 
cocones and colimits in categories. 
For each decorated theory $\Tt$ with pure subcategory $\Tt_0$, 
let $\Tt_A=F_{\expan}\Tt$ and $\tt_A\colon \Tt_A\to\Tt$ 
be as in section~\ref{subsec:gene}, 
and let $\Tt_a$ and $j_A\colon \Tt_A\to\Tt_a$ be as above. 
Let $j\colon \Tt\to\Tt_a$ be the morphism under $\Tt_0$ 
which maps each type $X$ to $X$ 
and each term $f\colon X\to Y$ to $f'\circ(a\times\id_X)\colon X\to Y$.
Let $t\colon j\circ\tt_A \To j_A$ be the natural transformation under $\Tt_0$ 
such that $t_A=a\colon \uno\to A$.
Then the following diagram is a \emph{lax cocone with base $\tt_A$} 
in the 2-category $\Tt_0\cosl\catT_{\eq}$, for short it is denoted $(\Tt_a,j_A,j,t)$,
and it is called \emph{the lax colimit associated to}~$\Tt$ because of lemma~\ref{lemm:passing}.
$$ \xymatrix@R=.8pc{
  \Tt_A \ar[dr]^{j_A} \ar[dd]_{\tt_A} & \\
  \ar@{}[r]|(.4){\trnat}|(.25){t}  & \Tt_a  \\ 
  \Tt \ar[ur]_{j} & \\ 
} $$

\begin{lemm}
\label{lemm:passing}
Let $\Tt$  be a decorated theory with pure subcategory $\Tt_0$.
The lax cocone $(\Tt_a,j_A,j,t)$ with base $\tt_A$ defined above 
is a lax colimit in the 2-category of theories under $\Tt_0$.
\end{lemm}

\begin{proof}
This means that the given lax cocone 
is initial among the lax cocones with base $\tt_A$ in $\Tt_0\cosl\Tt$,
in the following sense. 
For every lax cocone $(\Tt'_a,j'_A,j',t')$ with base $\tt_A$ under $\Tt_0$
there is a unique morphism $ h\colon \Tt_a\to\Tt'_a$ such that 
$ h\circ j_A=j'_A$, $ h\circ j=j'$ and $ h\circ t = t'$,
it is defined from the pushout in definition~\ref{defi:adding} 
by $h\circ j_A=j'_A$, so that $h(A)=A$, 
and $h\circ \gamma_a(a)=t'_A:\uno\to A$. 
\end{proof}

For instance, given $\Tt$, $\tt_A\colon  \Tt_A\to \Tt$, $\id_{\Tt}\colon  \Tt\to \Tt$ 
and $\id_{\tt_A}\colon \tt_A \To \tt_A$,
then $\tt_a$ is the unique morphism such that 
$\tt_a\circ j_A = \tt_A$, $\tt_a\circ j = \id_{\Tt}$ and $\tt_a\circ t=\id_{\tt_A}$.

Let $\Tt$ be a decorated theory with pure subtheory $\Tt_0$
and let $(\Tt_a,j_A,j,t)$ be its associated lax colimit, with base $\tt_A\colon \Tt_A\to \Tt$.
Let $M_A$ be a model of $\Tt_A$
and $M_0$ its restriction to $\Tt_0$, 
and let $\{ (M,m) \mid  m\colon  {\tt_A}^\md M \to M_A  \}|_{M_0}$
(where as before ${\tt_A}^\md M=M\circ\tt_A$) 
denote the set of pairs $(M,m)$ with $M$ a model of $\Tt$ over $M_0$
and $m$ a morphism of models of $\Tt_A$ over $M_0$. 
A consequence of the lax colimit property 
is that the function which maps each model $M_a$ of $\Tt_a$ over $M_A$
to the pair $(j^\md M_a, t^\md M_a)=(M_a\circ j,M_a\circ t)$ 
defines a bijection: 
\begin{equation}
\label{eq:passing}
 \Mod(\Tt_a)|_{M_A}  \iso
 \{ (M,m) \mid  m\colon  {\tt_A}^\md M \to M_A  \}|_{M_0} \;.
\end{equation}

The bijections~\ref{eq:adding} and~\ref{eq:passing} provide the next result,
which does not involve $\Tt_a$. 

\begin{prop} 
\label{prop:passing}
Let $\Tt$ be a decorated theory with pure subtheory $\Tt_0$
and let $\Tt_A=F_{\expan}\Tt$ and $\tt_A\colon \Tt_A\to \Tt$.
Then for each model $M_A$ of $\Tt_A$, 
with $M_0$ denoting the restriction of $M_A$ to $\Tt_0$, 
the function which maps each element $\alpha\in M_A(A)$
to the pair $(M,m)$, 
where $M$ is the model of $\Tt$ such that $M(f)=M_A(f')(\alpha,-)$ 
and where $m:{\tt_A}^\md M \to M_A$ is the morphism of models of $\Tt_A$ 
such that $m_A:M(\uno)\to M_A(A)$ is the constant function $\star\to\alpha$, 
defines a bijection: 
\begin{equation}
\label{eq:adding-passing}
M_A(A)  \iso
 \{ (M,m) \mid  m\colon  {\tt_A}^\md M \to M_A  \}|_{M_0} \;.
\end{equation}
\end{prop}

As an immediate consequence, 
we get the \emph{exact parameterization} property from \cite{LPR03}. 

\begin{coro} 
\label{coro:exact}
Let $\Tt$ be a decorated theory with pure subcategory $\Tt_0$,
and let $\Tt_A=F_{\expan}\Tt$.
Let $M_0$ be a model  of $\Tt_0$
and $M_A$ a terminal model of $\Tt_A$ over $M_0$. 
Then there is a bijection: 
\begin{equation}
\label{eq:exact}
M_A(A)  \iso \Mod(\Tt)|_{M_0} 
\end{equation}
which maps each $\alpha\in M_A(A)$ to the model 
$M_{A,\alpha}$ of $\Tt$ defined by 
$M_{A,\alpha}(X)=M_0(X)$ for each type $X$
and $M_{A,\alpha}(f)=M_A(f')(\alpha,-)$ for each term $f$,
so that $M_{A,\alpha}(f)=M_A(f)$ for each pure term $f$.
\end{coro} 

The existence of a terminal model of $\Tt_A$ over $M_0$ 
is a consequence of \cite{Ru00} and \cite{HR95}.
Corollary~\ref{coro:exact} corresponds to the way algebraic structures are implemented 
in the systems Kenzo/EAT.
In these systems 
the parameter set is encoded by means of a record of Common Lisp functions, 
which has a field for each operation in the algebraic structure to be implemented. 
The pure terms correspond to functions which can be obtained 
from the fixed data and do not require an explicit storage. 
Then, each particular instance of the record gives rise to an algebraic structure.


Clearly the construction of $\gamma_a$ from $\gamma_A$ 
is a functor, which is left adjoint to the functor 
which simply forgets that the constant $a$ is distinguished.
So, by composing this adjunction 
with the adjunction $F_{\expan}\dashv G_{\expan}$ from theorem~\ref{theo:gene} 
we get an adjunction $F'_{\expan}\dashv G'_{\expan}$
where $F'_{\expan}$ maps each decorated theory $\Tt$ to $\Tt_a$, as defined above: 
  $$ \xymatrix@C=4pc{ \catT_{\dec} \ar@/^/[r]^{F'_{\expan}} \ar@{}[r]|{\bot} & 
  \catT_a \ar@/^/[l]^{G'_{\expan}} } $$
Let $U'\colon \catT_a\to\catT_{\eq}$ be the functor 
which simply forgets that the type $A$ and the constant $a$ are distinguished. 
Then the functor $U' \circ F'_{\expan} \colon \catT_{\dec}\to\catT_{\eq}$ 
maps the decorated theory $\Tt$ to the equational theory $\Tt_a$.
  $$ \xymatrix@C=3pc{ 
  \catT_{\dec} \ar[r]^{F'_{\expan}} &  \catT_a  \ar[r]^{U'} &  \catT_{\eq} \\ } $$ 
The morphism of theories $j\colon \Tt\to\Tt_a$ depends on the decorated theory $\Tt$,
let us denote it $j=J_{\Tt}$.
Let $H\colon \catT_{\dec}\to\catT_{\eq}$ be the functor which maps 
each decorated theory $\Tt$ to the equational theory $\Tt$.
The next result is easy to check.

\begin{theo}
\label{theo:passing}
The morphisms of theories $J_{\Tt}\colon \Tt\to\Tt_a$ form the components of 
a natural transformation $J\colon H \To U'\circ F'_{\expan}\colon \catT_{\dec}\to\catT_{\eq}$.
  $$ \xymatrix@C=3pc{ 
  \catT_{\dec} \ar[r]^{F'_{\expan}}  \ar@/_4ex/[rr]_{H}^(.4){\Uparrow}^(.45){J}  & 
  \catT_a  \ar[r]^{U'} &  \catT_{\eq} \\ } $$ 
\end{theo}

\begin{defi}
\label{defi:passing}
The natural transformation $J\colon H \To U'\circ F'_{\expan}\colon \catT_{\dec}\to\catT_{\eq}$
in theorem~\ref{theo:passing} is called the \emph{parameter passing natural transformation}.
\end{defi}


\begin{exam}
\label{exam:adding}
Starting from $\Tt_{\oper}$ and $\Tt_{\oper,0}$ as in example~\ref{exam:term}, 
the pushouts of theories from proposition~\ref{prop:general} 
and definition~\ref{defi:adding} are respectively: 

$$ \begin{array}{|c|c|c|c|c|c|c|}
\cline{1-1} \cline{3-3} \cline{5-5} \cline{7-7} 
  
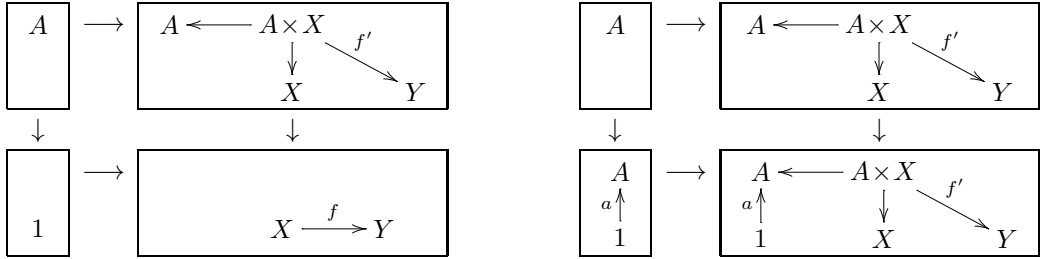
\xymatrix{A \\} &  
  \longrightarrow & 
  \xymatrix@R=1pc{A & A\stimes X \ar[l]\ar[d]\ar[rd]^{f'} & \\ & X & Y \\} &
  \qquad  \qquad & 
  \xymatrix@R=1pc{A \\} &  
  \longrightarrow & 
  \xymatrix@R=1pc{A & A\stimes X \ar[l]\ar[d]\ar[rd]^{f'} & \\ & X & Y \\} \\
\cline{1-1} \cline{3-3} \cline{5-5} \cline{7-7} 
  \col{\downarrow} & 
  \col{} & 
  \col{\downarrow} & 
  \col{} & 
  \col{\downarrow} & 
  \col{} & 
  \col{\downarrow} \\
\cline{1-1} \cline{3-3} \cline{5-5} \cline{7-7} 
  \xymatrix@R=1pc{ \\ \uno  \\ } &  
  \longrightarrow & 
  \xymatrix@R=1pc{ \\ & X\ar[r]^{f} &Y \\ } & & 
  \xymatrix@R=1pc{A \\ \uno \ar[u]^{a} \\ } &  
  \longrightarrow & 
  \xymatrix@R=1pc{ A & A\stimes X \ar[l]\ar[d]\ar[rd]^{f'} & \\ \uno \ar[u]^{a} & X & Y \\ } \\ 
\cline{1-1} \cline{3-3} \cline{5-5} \cline{7-7} 
\end{array}  $$

We have seen in example~\ref{exam:term} 
two other presentations of the vertex $\Tt_{\oper,a}$ of the second pushout,
with $f''=f'\circ(a\times\id_X):X\to Y$.
For each decorated theory $\Tt$, 
the morphism of equational theories $j_{\oper}=J_{\Tt_{\oper}}:\Tt\to\Tt_a$ 
maps $f$ to $f''$, as in example~\ref{exam:term}.

A model $M_0$ of $\Tt_{\oper,0}$ is simply made of two sets 
$\setX=M_0(X)$ and $\setY=M_0(Y)$.
On the one hand, a model of $\Tt$ over $M_0$ is characterized 
by a function $\varphi\colon \setX\to\setY$.
On the other hand, 
the terminal model $M_A$ of $\Tt_{\oper,A}$ over $M_0$ 
is such that $M_A(A)=\setY^\setX$ and 
$M_A(f')\colon \setY^{\setX} \times \setX \to \setY$ is the application. 
The bijection $\Mod(\Tt)|_{M_0}\iso M_A(A)$ then 
corresponds to the currying bijection $\varphi\mapsto\curry{\varphi}$. 
\end{exam}

\begin{exam}
\label{exam:dm-terminal}
Let $\Tt_{\dm}$ be the theory for differential monoids from example~\ref{exam:dm},
with the pure subtheory $\Tt_{\dm,0}=\Tt_{\mon}$ of monoids.
They generate the parameterized theory $\Tt_{\dm,A}$ as in example~\ref{exam:dm}.
Let $M_0$ be some fixed monoid 
and $M_A$ any model of $\Tt_{\dm,A}$ over $M_0$,
then each element of $M_A(A)$ corresponds to a
differential structure on the monoid $M_0$.
If in addition $M_A$ is the terminal model of $\Tt_{\dm,A}$ over $M_0$, 
then this correspondence is bijective.
\end{exam}

\begin{exam}
\label{exam:state}
When dealing with an imperative language, 
the states for the memory are endowed with an operation $\lup$
for observing the state and an operation $\upd$ for modifying it. 
There are two points of view on this situation: either the state is hidden,
or it is explicit.
Let us check that the parameterization process allows to generate 
the theory with explicit state from the theory with hidden state.

First, let us focus on observation: 
the theory $\Tt_{\st}$ is made of two types $L$ and $Z$
(for locations and integers, respectively) and a term $v\colon L\to Z$
for observing the values of the variables.
The pure subtheory $\Tt_{\st,0}$ is made of $L$ and $Z$.
We choose a model $M_0$ of $\Tt_{\st,0}$ made of a countable set 
of locations (or addresses, or ``variables'') $\bL=M_0(L)$ 
and of the set of integers $\bZ=M_0(Z)$.
Let $\bA=\bZ^\bL$, then as in example~\ref{exam:adding} 
the terminal model $M_A$ of $\Tt_{\st,A}$ over $M_0$ 
is such that $M_A(A)=\bA$ and $M_{\st,A}(v')\colon \bA\times\bL\to\bZ$ 
is the application, denoted $\lup$. 
The terminal model $M_A$ does correspond to an ``optimal'' implementation 
of the state.

Now, let us look at another model $N_A$ of $\Tt_{\st,A}$ over $M_0$,
defined as follows: 
$N_A(A)=\bA\times \bL\times \bZ$ and 
$N_A(v')\colon \bA\times \bL\times \bZ\times \bL\to\bZ$
maps $(p,x,n,y)$ to $n$ if $x=y$ and to $\lup(p,y)$ otherwise.
The terminality property of $M_A$ ensures that there is a unique 
function $\upd\colon \bA\times \bL\times \bZ \to \bA$ such that 
$\lup(\upd(p,x,n),y)$ is $n$ if $x=y$ and $\lup(p,y)$ otherwise.
So, the updating operation $\upd$ is defined coinductively from the observation operation $\lup$. 
\end{exam}

\section{Free functors}
\label{sec:free}

In this section some basic facts about limit sketches and their associated adjunction
are mentioned, 
and it is proved that the parameterization functor $F_{\expan}$ 
from section~\ref{subsec:gene} is a free functor,
in the sense that it is the left  adjoint 
associated to a morphism of limits sketches. 

\subsection{Limit sketches}
\label{subsec:sketch}

It is quite usual to define a \emph{free} functor 
as the left adjoint of a forgetful functor,
but there is no unique definition of a forgetful functor.
In this section forgetful functors are defined 
from morphisms of limit sketches, they are not always faithful. 

There are several definitions of limit sketches
(also called projective sketches) in the litterature,
see for instance \cite{CL84} or \cite{BW99}. 
These definitions are different but all of them serve the same purpose: 
each limit sketch generates a category with limits,
so that limit sketches generalize equational specifications 
in allowing some interdependence between the variables.
In this paper, limit sketches are used at the meta level,
in order to describe each category of theories or specifications 
as the category of realizations (or models) of a limit sketch.
While a category with limits is a graph with identities, composition, 
limit cones and tuples, satisfying a bunch of axioms, 
a limit sketch is a graph with \emph{potential} identities, composition, 
limit cones and tuples, which are not required to satisfy any axiom.
Potential limit cones, or simply \emph{potential limits}, 
may also be called \emph{specified limits} or \emph{distinguished cones}.

\begin{defi}
\label{defi:sketch}
A \emph{limit sketch} is a graph where 
some points $X$ 
  have an associated potential identity arrow $\id_X\colon X\to X$,
some pairs of consecutive arrows $f\colon X\to Y$, $g\colon Y\to Z$
  have an associated potential composed arrow $g\circ f\colon X\to Z$,
some diagrams $\Delta$ 
  have an associated potential limit, which is a cone with base $\Delta$, 
and when there is a potential limit with base $\Delta$ then some cones with base $\Delta$
  have an associated potential tuple, which is a morphism of cones with base $\Delta$
  from the given cone to the potential limit cone. 
A morphism of limit sketches is a morphism of graphs which preserves the 
potential features. This yields the category of limit sketches.
\end{defi}

Whenever this definition is restricted to potential limits with a finite discrete base
(called potential finite products), 
we get the category of \emph{finite product sketches}:
this is the category $\catS_{\eq}$ of equational specifications, from section~\ref{subsec:equa}.

\begin{defi}
\label{defi:realization}
Given a limit sketch $\skE$ and a category $\catC$,
a \emph{realization} (or \emph{loose model}) of $\skE$ with values in $\catC$
is a graph homomorphism which maps the potential features of $\skE$
to real features of $\catC$.
A morphism of realizations is (an obvious generalization of) a natural transformation.
This gives rise to the category $\Rea(\skE,\catC)$
of realizations of $\skE$ with values in $\catC$.
By default, $\catC$ is the category of sets.
\end{defi}

By default, $\catC$ is the category of sets.
A category is called \emph{locally presentable} 
if it is equivalent to the category of set-valued realizations 
of a limit sketch $\skE$; then $\skE$ is called a limit sketch \emph{for} this category.

Let $\olskE$ denote the category generated by $\skE$
such that every potential potential feature of $\skE$ becomes a real feature of $\olskE$. 
The \emph{Yoneda contravariant realization} $\funY_{\skE}$ of $\skE$ 
is the contravariant realization of $\skE$ with values in $\Rea(\skE)$
such that $\funY_{\skE}(E)=\Hom_{\olskE}(E,-)$
for every point or arrow $E$ in $\skE$. 
Then for each theory $\Tt$ and each point $E$ in $\skE$,
the set $\Tt(E)$ is in bijection with $\Hom_{\Rea(\skE)}(\funY_{\skE}(E),\Tt)$.
The Yoneda contravariant realization is injective on objects and faithful. 
In addition it is \emph{dense}:
although $\Rea(\skE)$ may be ``much larger'' than $\skE$, 
every realization of $\skE$ is the vertex of a colimit 
with its base in the image of $\funY_{\skE}$.

Let $\ske\colon \skE_1\to\skE_2$ 
be a morphism of limit sketches and 
$G_{\ske}\colon \Rea(\skE_2)\to\Rea(\skE_1)$ the precomposition with $\ske$. 
A fundamental result due to Ehresmann states that there is an adjunction,
that will be called the adjunction \emph{associated with} $\ske$: 
  $$ \xymatrix@C=4pc{ \Rea(\skE_1) \ar@/^/[r]^{F_{\ske}} \ar@{}[r]|{\bot} & 
  \Rea(\skE_2) \ar@/^/[l]^{G_{\ske}} } $$
Moreover, the functor $F_{\ske}$ \emph{(contravariantly) extends} $\ske$
via the Yoneda contravariant realizations,
in the sense that there is an isomorphism:
  $$F_{\ske} \circ \funY_{\skE_1} \iso \funY_{\skE_2} \circ \ske \;.$$
Our definition of forgetful and free functors relies on this adjunction.

\begin{defi}
\label{defi:free}
A \emph{forgetful} functor is a functor of the form
$G=-\circ\ske\colon \Rea(\skE_2)\to\Rea(\skE_1)$
for a morphism of limit sketches $\ske\colon \skE_1\to\skE_2$.
A \emph{free} functor is a left adjoint to a forgetful functor
(as every adjoint functor, it is unique up to a natural isomorphism).
\end{defi}

\begin{rema}
\label{rema:free}
It is easy to describe the forgetful functor $G_{\ske}$, using its definition: 
for each realization $R_2$ of $\skE_2$, 
the realization $R_1=G_{\ske}(R_2)$ of $\skE_1$ is such that $R_1(E)=R_2(\ske(E))$
for every point or arrow $E$ in $\skE_1$. 
It is also quite easy to describe the left adjoint functor $F_{\ske}$,
using the fact that $F_{\ske}$ extends $\ske$:
let $R_1$ be a realization of $\skE_1$ and $R_2=F_{\ske}(R_1)$,
if $R_1=\funY_{\skE_1}(E)$ for some point $E$ in $\skE_1$ 
then $R_2=\funY_{\skE_2}(\ske(E))$,
and the general case follows thanks to the density of $\funY_{\skE_1}$
and to the fact that $F_{\ske}$ preserves colimits (since it is a left adjoint).
\end{rema}

\subsection{A limit sketch for equational theories}
\label{subsec:sk-equa}

The construction of various ``sketches of categories'' and ``sketches of sketches'' 
is a classical exercise about sketches \cite{CL84,CL88,BW99}. 
Here we build (a significant part of) a limit sketch $\skE_{\eq}$ 
for the category $\catT_{\eq}$ of equational theories,
i.e., for the category of categories with chosen products.

\subsubsection*{$\bullet$ Graphs}

Let us start from the following limit sketch $\skE_{\gr}$ for the category of graphs,
simply made of two points $\Type$ and $\Term$ (for types and terms)
and two arrows $\dom$ and $\codom$ (for domain and codomain):
$$
  \xymatrix@C=4pc{
    \;\Type\; & \;\Term\; \ar[l]_{\dom} \ar@<1ex>[l]^{\codom} \\
    }
$$

The image of $\skE_{\gr}$ by its Yoneda contravariant realization 
is the following diagram of graphs:
$$ \begin{array}{|c|c|c|}
  \cline{1-1}\cline{3-3}
  \xymatrix{
    X \\}  &
  \xymatrix@C=3pc{
    \mbox{} \ar[r]^{X\mapsto X}  \ar@<-2ex>[r]_{X\mapsto Y} & \mbox{} \\}  &
  \xymatrix{
    X \ar[r]^{f} & Y \\}  \\
  \cline{1-1}\cline{3-3}
  \end{array}$$

\subsubsection*{$\bullet$ Categories}

First, let us build a limit sketch $\skE'_{\gr}$
by adding to $\skE_{\gr}$ a point $\Cons$ for consecutive terms,
as the vertex of the following potential limit, where the projections 
$\fst$ and $\snd$ stand for the first and second component  
of a pair of consecutive terms and $\midd$ stands for its ``middle type'' 
(codomain of the first component and domain of the second one):
$$
\xymatrix@=1.5pc{
  & \Cons \ar[dl]_{\fst} \ar[dr]^{\snd} \ar[d]|{\midd} & \\
  \Term \ar[r]_{\codom} & \Type & \Term \ar[l]^{\dom} \\
  } $$
hence the equations 
$ \codom\circ\fst = \midd \,,\; \dom\circ\snd = \midd $ hold, 
so that $\midd$ may be omitted. 
Adding such a potential limit, with new vertex and projections over a known base,
is an equivalence of limit sketches:
the realizations of $\skE'_{\gr}$ are still the graphs.
Now, a limit sketch $\skE_{\cat}$ for categories is obtained by adding to $\skE'_{\gr}$
two arrows $\selid$ for the selection of identities and $\comp$ for the composition 
and several equations: 
$$
  \xymatrix@C=4pc{
    \;\Type\; \ar@<1ex>@/^/[r]^{\selid} & 
    \;\Term\; \ar[l]_{\dom} \ar@<1ex>[l]^{\codom} & 
    \;\Cons\; \ar[l]_{\fst}  \ar@<1ex>[l]^{\snd} \ar@<-1ex>@/_/[l]_{\comp} \\
    } $$
$$ 
  \dom\circ\selid = \ttid_{\Type} \,,\;
  \codom\circ\selid = \ttid_{\Type} \,,\;
  \dom\circ\comp = \dom\circ\fst \,,\;
  \codom\circ\comp = \codom\circ\snd \,,
$$
and with the equations which ensure that the three axioms of categories are satisfied.
The image of this part of $\skE_{\cat}$ by its Yoneda contravariant realization 
is the following diagram of categories: 
$$ \begin{array}{|c|c|c|c|c|}
  \cline{1-1}\cline{3-3} \cline{5-5}
   
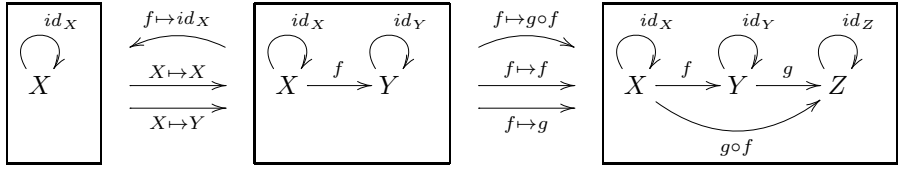
\xymatrix{
    X \ar@(lu,ru)^(.6){\id_X} \\}  & 
  \xymatrix@C=3pc{
    \mbox{}  \ar[r]^{X\mapsto X} \ar@<-2ex>[r]_{X\mapsto Y} &
    \mbox{} \ar@<-3ex>@/_/[l]_{f\mapsto\id_X} \\ } &
  \xymatrix{
    X \ar@(lu,ru)^(.6){\id_X} \ar[r]^{f} & 
    Y \ar@(lu,ru)^(.6){\id_Y} \\}  & 
  \xymatrix@C=3pc{
    \mbox{} \ar[r]^{f\mapsto f} \ar@<-2ex>[r]_{f\mapsto g} 
       \ar@<3ex>@/^/[r]^{f\mapsto g\circ f} & 
    \mbox{} \\} &
  \xymatrix{
    X \ar@(lu,ru)^(.6){\id_X} \ar[r]^{f} \ar@/_4ex/[rr]_{g\circ f} & 
    Y \ar@(lu,ru)^(.6){\id_Y}  \ar[r]^{g} & 
    Z \ar@(lu,ru)^(.6){\id_Z} \\}  \\
  \cline{1-1}\cline{3-3} \cline{5-5}
  \end{array}$$

\subsubsection*{$\bullet$ Theories}

We build a limit sketch $\skE'_{\cat}$
by adding to $\skE_{\cat}$ for each $n\in\bN$ the following potential limits, 
with vertex $\nType$ for $n$-tuples of types and $\nCone$ for $n$-ary discrete cones:
$$
  
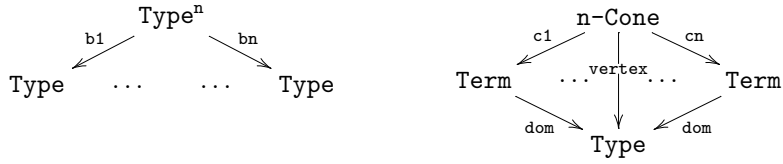
\xymatrix@=1pc{
  & \nType \ar[dl]_{\prbone} \ar[dr]^{\prbn} & \\
  \Type & \dots \qquad \dots & \Type \\
  } \qquad\qquad 
  \xymatrix@=1pc{
  & \nCone \ar[dl]_{\prcone} \ar[dr]^{\prcn} \ar[dd]|(.4){\vertex} & \\
  \Term \ar[rd]_{\dom} & \dots \qquad \dots & \Term \ar[ld]^{\dom} \\
   & \Type  &  \\
  }
$$ 
The arrow $\vertex$ may be omitted when $n>0$.
When $n=0$, the potential limits mean that 
$\zType$ is a unit type (also denoted $\Unit$) 
and $\zCone$ is isomorphic to $\Type$. 
We also add the tuple 
$\nbase=\tuple{\codom\circ\prcone, \dots, \codom\circ\prcn}$
which maps each cone to its base:
  $$
  \xymatrix@C=4pc{
 \nType & \nCone \ar[l]_{\nbase} 
    }
$$
The realizations of $\skE'_{\cat}$ are still the categories.
Now, a limit sketch $\skE_{\thry}$ for equational theories is obtained by adding to $\skE'_{\cat}$
the following features, for each $n\in\bN$. 
First an arrow $\nprod\colon \nType\to\nCone$ 
together with the equation $\nbase\circ\nprod=\ttid_{\nType}$,
for building the product cone of each family of $n$ types. 
Then for building tuples, an arrow $\ntuple\colon \nCone\to\Term$ 
together with the equations $\dom\circ\ntuple=\vertex$
and $\codom\circ\ntuple=\vertex\circ\nprod\circ\nbase$
and with several additional equations for ensuring that the universal property of a product
is satisfied.
So, here is a relevant part of this limit sketch $\skE_{\thry}$ for theories 
(equations are omitted, and only one arity $n$ is represented): 

$$
  
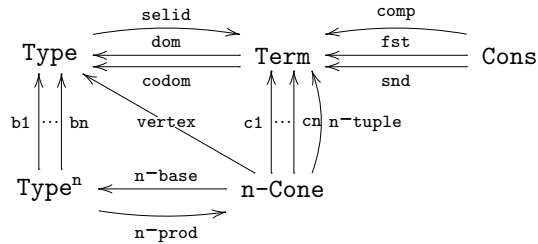
\xymatrix@C=4pc@R=3pc{
    \;\Type\; \ar@<1ex>@/^/[r]^{\selid} & 
    \;\Term\; \ar[l]_{\dom} \ar@<1ex>[l]^{\codom} & 
    \;\Cons\; \ar[l]_{\fst}  \ar@<1ex>[l]^{\snd} \ar@<-1ex>@/_/[l]_{\comp} \\
    \;\nType\; \ar@<1ex>[u]^{\prbone} \ar@{}[u]|{\dots} \ar@<-1ex>[u]_{\prbn} 
      \ar@<-1ex>@/_/[r]_{\nprod} & 
    \;\nCone\; \ar@<1ex>[u]^{\prcone} \ar@{}[u]|{\dots} \ar@<-1ex>[u]_{\prcn}
       \ar@<-2ex>@/_/[u]_{\ntuple} \ar[l]_{\nbase} \ar[ul]|{\vertex} &
    } $$

Let us focus on the following part of $\skE_{\thry}$:

$$
  \xymatrix@C=4pc{
    \;\bType\; \ar@<-1ex>@/_/[r]_{\bprod} & 
    \;\bCone\; \ar[l]_{\bbase}  \ar@/_/[r]_{\btuple} &
    \;\Term\; \\
    }
$$

and its image by the Yoneda contravariant realization (only presentations are given): 

$$ \begin{array}{|c|c|c|c|c|}
  \cline{1-1}\cline{3-3}\cline{5-5}
  
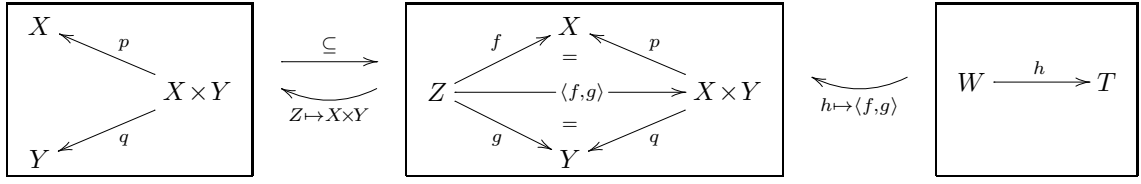
\xymatrix@C=3pc@R=1pc{ X & \\ 
  & X\stimes Y \ar[lu]_{p} \ar[ld]^{q} \\ 
  Y & \\ } &  
  \xymatrix@C=3pc@R=1pc{ \\ 
    \mbox{} \ar@<1ex>[r]^{\subseteq} &
    \mbox{} \ar@<1ex>@/^/[l]^{Z\mapsto X\stimes Y} \\ } &
  \xymatrix@C=3pc@R=1pc{ & X & \\ 
  Z \ar[ru]^{f} \ar[rd]_{g} \ar[rr]|{\,\tuple{f,g}\,} & \ar@{}[u]|{=}\ar@{}[d]|{=} & 
  X\stimes Y \ar[lu]_{p} \ar[ld]^{q} \\ & Y & \\ } &
  \xymatrix@C=3pc@R=1pc{ \\ 
    \mbox{}  &
    \mbox{} \ar@/^/[l]^{h\mapsto \tuple{f,g}} \\ } &
  \xymatrix@C=3pc@R=1pc{ \mbox{} \\ W \ar[r]^{h} & T \\ } \\  
  \cline{1-1}\cline{3-3}\cline{5-5}
  \end{array}$$

\subsection{The parameterization process is a free functor} 
\label{subsec:sk-free}

\subsubsection*{$\bullet$ Parameterized theories}

A limit sketch $\skE_A$ for parameterized theories is obtained 
by adding to $\skE_{\thry}$ an arrow $\ttA \colon \Unit \to \Type$.

\subsubsection*{$\bullet$ Decorated theories}

A limit sketch $\skE_{\dec}$ for decorated theories comes with
a morphism $\ske_{\undec}\colon \skE_{\dec}\to\skE_{\thry}$ 
which forgets about the decorations
(``$\mathit{undec}$'' for ``undecoration'').
Here are two slightly different choices,
the first one is simpler but the second one better reflects the idea of decoration.

A limit sketch  $\skE_{\dec}$ for decorated theories is made of 
two related copies of $\skE_{\thry}$: 
one copy $\skE\pur$ for the pure features
and another copy $\skE\gen$ for the general features,
together with a monomorphic transition arrow $\ttt_{\ttE} \colon \ttE\pur \to \ttE\gen$ 
for each point $\ttE$ in $\skE_{\thry}$ with $\ttt_{\Type}$ an identity 
and with the transition equations 
$\ttt_{\ttE'}\circ\tte\pur=\tte\gen\circ\ttt_{\ttE}$ 
for each arrow $\tte\colon \ttE\to\ttE'$ in $\skE_{\thry}$.
The morphism $\ske_{\undec}$ maps both copies $\skE\pur$ and $\skE\gen$ to $\skE$.

Another limit sketch for decorated theories, still denoted $\skE_{\dec}$,
is the \emph{sketch of elements} (similar to the more usual \emph{category of elements}) 
of a model $\catD$ of $\skE_{\thry}$ with values in $\catT_{\eq}$,   
then the morphism $\ske_{\undec}$ is provided by the construction.
This model $\catD$ formalizes the fact that 
identities and projections are always pure
while the composition or pairing of pure terms is pure.
Precisely, the theory $\catD(\Type)$ is generated by one type $D$
and the theory $\catD(\Term)$ by two types $p$, $g$ 
(for ``pure'' and ``general'', respectively) and a monomorphism $p\to g$
(for ``every pure term can be seen as a general term'').
As for the functors, $\catD(\selid)$ maps $D$ to $p$,  
$\catD(\nprod)$ maps $\tuple{D,\dots,D}$ to $p$,
while $\catD(\comp)$ maps
$\tuple{p,p}$ to $p$ and $\tuple{p,g},\tuple{g,p},\tuple{g,g}$ to $g$
and $\catD(\ntuple)$ maps
$\tuple{p,\dots,p}$ to $p$ and everything else to $g$.
The resulting sketch of elements $\skE_{\dec}$ is made of 
one point $\Type\mathtt{.D}$ over the point $\Type$ of $\skE_{\thry}$,
two points $\Term\pur$ and $\Term\gen$ over the point $\Term$ of $\skE_{\thry}$,
four points over $\Cons$, $2^n$ over $\nType$ and $\nCone$,
a monomorphic arrow $\Term\pur \to \Term\gen$, and so on.

\subsubsection*{$\bullet$ From decorated theories to parameterized theories}

Let us consider the functors $F_{\expan}\colon \catT_{\dec}\to\catT_A$
and $G_{\expan}\colon \catT_A\to\catT_{\dec}$ from section~\ref{subsec:gene}. 
We can now prove that $G_{\expan}$ is a forgetful functor 
and $F_{\expan}$ is its associated free functor, 
in the sense of definition~\ref{defi:free}.

\begin{theo}
\label{theo:free}
There is a morphism of limit sketches $\ske_{\expan}\colon  \skE_{\dec} \to \skE_A$ 
such that  the associated adjunction is $F_{\expan}\dashv G_{\expan}$
from section~\ref{subsec:gene}. 
\end{theo}

\begin{proof}
In section~\ref{subsec:gene} the functor $F_{\expan}$ has been defined 
on $\funY(\skE_{\dec})$ (see figure~\ref{fig:F-elem}).
Since $F_{\expan}$ extends $\ske_{\expan}$ via the Yoneda contravariant realizations,
this provides the definition of a unique morphism $\ske_{\expan}$ 
with associated left adjoint $F_{\expan}$. 
For instance, the point $\Term\pur$ is mapped to $\Term$
and the point $\Term\gen$ to the point of $\skE_A$ 
characterized by the fact 
that its image by Yoneda is presented by $X$, $Y$ and $f'\colon A\times X\to Y$.
Then it is easy to check that the precomposition with $\ske_{\expan}$ 
is the functor $G_{\expan}$.
\end{proof}

\subsubsection*{$\bullet$ A span of limit sketches}

Altogether, the following span of limit sketches provides a framework
for the process that starts from an equational theory,
choose the pure terms, and forms the corresponding parameterized theory:
$$ \xymatrix@C=4pc { 
\skE_{\eq} & \skE_{\dec} \ar[l]_{\ske_{\undec}} \ar[r]^{\ske_{\expan}} & \skE_A \\
\\ } $$

\subsection{A limit sketch for equational specifications} 
\label{subsec:sk-spec}

In this section we build a limit sketch $\skE_{\spec}$ for equational specifications
from the limit sketch $\skE_{\thry}$ for theories,
thus providing another point of view on the elementary specifications
in section~\ref{subsec:equa}. 
This construction can be seen as an illustration of the 
factorization theorem in \cite{Du03}. 
A direct detailed construction of a limit sketch for equational specifications
can be found in appendix~\ref{subsec:dia-spec}.

In the part of $\skE_{\thry}$ shown in section~\ref{subsec:sk-equa}
there are four arrows that are neither in $\skE_{\gr}$ nor
projections in a potential limit: 
$\selid$, $\comp$, $\nprod$, $\ntuple$.
These arrows stand for features that are always defined in a theory 
but only partially defined in a specification. 
So, $\skE_{\spec}$ is obtained by replacing each of these arrows 
$\tte\colon \ttE_1\to\ttE_2$ by a span: 
  $$\xymatrix@C=3pc{
     \ttE_1 & \;\;\ttE'_1\;\;  \ar@{>->}[l]_{\tte'_1}  \ar[r]^{\tte'} & \ttE_2 \\
  }$$
where the arrow ``$\rightarrowtail$'' stands for a potential monomorphism 
(which can be expressed as a potential limit).
So, here is (a relevant part of) $\skE_{\spec}$:
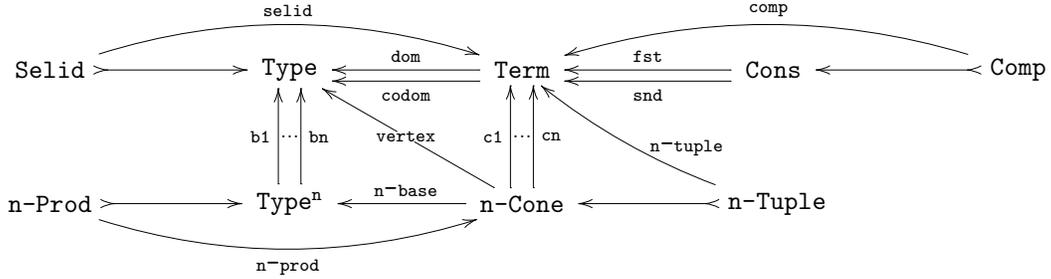
$$
  \xymatrix@C=4pc@R=3pc{
    \;\Selid\;\; \ar@{>->}[r] \ar@/^4ex/[rr]^{\selid} & 
    \;\Type\; & 
    \;\Term\; \ar[l]_{\dom} \ar@<1ex>[l]^{\codom} & 
    \;\Cons\; \ar[l]_{\fst}  \ar@<1ex>[l]^{\snd} &
    \;\;\Comp\; \ar@{>->}[l] \ar@/_4ex/[ll]_{\comp} \\
    \;\nProd\;\; \ar@{>->}[r] \ar@/_4ex/[rr]_{\nprod} & 
    \;\nType\; \ar@<1ex>[u]^{\prbone} \ar@{}[u]|{\dots} \ar@<-1ex>[u]_{\prbn} & 
    \;\nCone\; \ar@<1ex>[u]^{\prcone} \ar@{}[u]|{\dots} \ar@<-1ex>[u]_{\prcn}
       \ar[l]_{\nbase} \ar[ul]|{\vertex} &
    \;\;\nTuple\; \ar@{>->}[l] \ar@/^/[ul]_(.4){\ntuple}  & \\
    } $$
The elementary specifications from section~\ref{subsec:equa} 
are the images by the Yoneda contravariant realization 
of the points in $\skE_{\spec}$ which are not vertices of potential limits, namely:
$\Type$, $\Term$, $\Selid$, $\Comp$, $\nProd$, $\nTuple$:
our notations are such that $\Ss_{\ttE}=\funY(\ttE)$ for each of these points $\ttE$. 

\section{Conclusion}

This paper provides a neat categorical formalization for 
the parameterization process in Kenzo and EAT.
Future work includes the generalization of this approach 
from equational theories to other families of theories, like distributive categories,
and to more general kinds of parameters, like data types.


\appendix 

\section{Diagrammatic logics}
\label{sec:dia}

In this paper we have introduced the equational logic in a categorical way,
considering equational theories as categories with chosen finite products.
An equational theory can be presented by an equational specification,
which means that this specification generates the theory.
In section~\ref{sec:free} we have outlined the construction
first of a limit sketch for the equational theories
and then of a limit sketch for the equational specifications.
This appendix provides a detailed description of these limit sketches,
with slightly more subtle definitions of equational theories and specifications,
which are better suited for formalizing equational proofs. 
In addition, as often in the framework of algebraic specifications 
(as for instance in \cite{LEW96} and in \cite{DDLR05})
we consider first the specifications, then we get the theories 
by using the inference rules of the equational logic.
Finally, the parameterization process is presented from this point of view. 
The framework of diagrammatic logics \cite{Du03,Du07} is well suited 
for dealing with ``usual'' logics like the equational logic
as well as with more ``unusual'' ones like the decorated equational logic,
and also for dealing with various morphisms of logics, 
for instance we will see that the parameterizing functor 
stems from a  morphism of logics.
This appendix can be seen as an introduction to diagrammatic logics, 
based on section~\ref{subsec:sketch} about limit sketches.

\subsection{Equational logic, revisited}
\label{subsec:dia-equ}

As in section~\ref{sec:defi},
instead of the algebraic definition of equational specifications given
for instance in~\cite{LEW96},
we define equational specifications from finite product sketches. 
In the main text we have defined equational specifications 
exactly as finite product sketches,
so that the equations become equalities of arrows:
this is all right for defining models but this makes every proof trivial.
In this appendix we give a more subtle definition of equational specifications
as finite product sketches \emph{with equations};
then definition~\ref{defi:equa-spec} is easily recovered 
by mapping equations to equalities. 
In spite of this minor difference, 
we use the same notations ($\catS_{\eq}$, $\catT_{\eq}$)
in this appendix as in the main part of the paper.

\begin{defi}
\label{defi:dia-spec}
An \emph{equational specification} is a limit sketch (definition~\ref{defi:sketch})
where all the potential limits are potential finite products, 
together with a set of pairs of parallel terms called the \emph{equations}
and denoted $t_1 \equiv t_2$.
A morphism of equational specifications
is a morphism of limit sketches which preserves the equations.
This yields the category of equational specifications $\catS_{\eq}$. 
\end{defi}

Similarly, in this appendix, the equations in an equational theory
need not be equalities. 
Roughly speaking, an equational theory is an equational specification
where the equations form an equivalence relation
and all the potential features become real up to equations.
For this reason the relation $\equiv$ is called a \emph{congruence}.
So, an equational theory is not a category, it is only a bicategory
(the congruence defines its 2-cells),
but it becomes a category with chosen finite products, 
as in definition~\ref{defi:equa-thry}, 
as soon as both members in each equation get identified.
Conversely, every category with chosen finite products can be seen as
an equational theory where the equations are the equalities.

\begin{defi}
\label{defi:dia-thry}
An \emph{equational theory} is an equational specification
where  
each type has a potential identity,
each pair of consecutive terms has a potential composition,
each list of types has a potential product,  
each list of terms with a common domain has a potential tuple,
and in addition the equations form a \emph{congruence},
which means that the relation $\equiv$ is an equivalence relation
compatible with composition 
and that the usual axioms for categories with products are satisfied up to $\equiv$. 
The \emph{category of equational theories}
$\catT_{\eq}$ is the full subcategory of $\catS_{\eq}$
with objects the equational theories. 
\end{defi}

It may be noted that the inclusion of $\catT_{\eq}$ in $\catS_{\eq}$ is faithful. 
In fact, for products and tuples, only the arities $n=2$ and $n=0$ will be considered: 
the general case may easily be guessed, 
or alternatively one can use the fact that all finite products may be recovered
from the binary products and a terminal object.
A set of \emph{inference rules} for the equational logic, 
for generating an equational theory from an equational specification, 
is presented in figure~\ref{fig:equ}.
When there is no ambiguity, we often omit ``equational'' and ``potential''.

\renewcommand{\arraystretch}{2} 
\begin{figure}[ht] 
$$\begin{array}{|l|c|}
\hline
\multicolumn{1}{|c|}{\textrm{name}} & \textrm{rules} \\ 
\hline
\hline
  \textrm{composition} &
    \frac{f:X\to Y \quad g:Y\to Z}{g\circ f:X\to Z} \\  
  \textrm{identity} &
    \frac{X}{\id_X:X\to X} \\
\hline
  \textrm{equivalence} &
    \frac{f}{f\equiv f}   \qquad
    \frac{f\equiv g}{g\equiv f}   \qquad
    \frac{f\equiv g \quad g\equiv h}{f\equiv h} \\
\hline
  \textrm{substitution} &
    \frac{f:X\to Y \quad g_1\equiv g_2:Y\to Z}{g_1\circ f\equiv g_2\circ f:X\to Z} \\
  \textrm{replacement} &
    \frac{f_1\equiv f_2:X\to Y \quad g:Y\to Z}{g\circ f_1\equiv g\circ f_2:X\to Z} \\
\hline
  \textrm{associativity} &
    \frac{f:X\to Y \quad g:Y\to Z \quad h:Z\to W}{(h\circ g)\circ f \equiv h\circ (g\circ f)} \\
  \textrm{unit rules} & 
    \frac{f:X\to Y}{f\circ\id_X\equiv f } \qquad
    \frac{f:X\to Y}{\id_Y\circ f \equiv f}  \\
\hline
  \textrm{binary product} &
    \frac{Y_1 \quad Y_2}
    {Y_1\times Y_2} \qquad
    \frac{Y_1 \quad Y_2}
    { \proj_1:Y_1\times Y_2\to Y_1} \qquad
    \frac{Y_1 \quad Y_2}
    {\proj_2:Y_1\times Y_2\to Y_2} \\
  \textrm{pairing} &  
     \quad  \frac{f_1:X\to Y_1 \quad f_2:X\to Y_2}
     {\tuple{f_1,f_2} :X\to Y_1\times Y_2} \qquad 
     \frac{f_1:X\to Y_1 \quad f_2:X\to Y_2}
     {\proj_1\circ \tuple{f_1,f_2} \equiv f_1} \qquad
     \frac{f_1:X\to Y_1 \quad f_2:X\to Y_2}
     {\proj_2\circ \tuple{f_1,f_2} \equiv f_2} \quad \\
  \textrm{pairing uniqueness} & 
     \frac{f_1:X\to Y_1\quad f_2:X\to Y_2 \quad f :X\to Y_1\times Y_2
       \quad \proj_1\circ f \equiv f_1  \quad \proj_2\circ f \equiv f_2  }
     {\tuple{f_1,f_2} \equiv f} \\
\hline
  \textrm{terminal type} &
    \frac{}{\;\uno\;} \\
  \textrm{collapsing} &
    \frac{X}{\tu_X:X\to \uno} \\
  \textrm{collapsing uniqueness} &
     \frac{f:X\to \uno}
     {\tu_X \equiv f}  \\
\hline
\end{array}$$
\caption{\label{fig:equ} The rules for equational logic}
\end{figure}
\renewcommand{\arraystretch}{1.3} 

The \emph{models} of a specification $\Ss$
with values in a theory $\Tt$ are defined as
the morphisms of specifications from $\Ss$ to $\Tt$.
In addition, the \emph{morphisms of models} of  $\Ss$
with values in $\Tt$ can be defined in the usual natural way, 
so that there is a \emph{category of models} $\Mod(\Ss,\Tt)$
of $\Ss$ with values in $\Tt$.
The  \emph{category of set-valued models} of  $\Ss$  is the 
category of models of $\Ss$ with values in the category of sets
seen as an equational theory, with the cartesian products as potential products
and the equalities of functions as equations.

Each equational specification in the algebraic sense $\Spec$ gives rise to an 
equational specification $\Ss$:
each sort of $\Spec$ becomes a type of $\Ss$,
each list of sorts $X_1,\dots,X_n$ of $\Spec$
becomes a type $X_1\times\dots\times X_n$ of $\Ss$,
each operation or term $f\colon X_1\dots X_n\to Y$ of $\Spec$ becomes a term
$f\colon X_1\times\dots\times X_n\to Y$ of $\Ss$, and each equation
$f_1\equiv f_2$ of $\Spec$ becomes an equation $f_1\equiv f_2$ of $\Ss$;
for this purpose, both terms $f_1$ and $f_2$ in $\Spec$
must be considered as terms in all the variables that appear in $f_1$ or in $f_2$,
as explained for instance in \cite{BW99}. 
Then, the category of models of $\Spec$ in the algebraic sense
is isomorphic to the category of set-valued models of $\Ss$.

\begin{exam}
\label{exam:dia-sg}
The equational specification $\Ss_{\sgp}$ for semigroups
can be represented as a graph with an equation:
  $$ \begin{array}{c|cc|}
  \cline{2-3}
  \Ss_{\sgp} = &
  \xymatrix{
    G^2 \ar[r]^{\prd} & G \\
    } &
  \xymatrix{
  \txt{$\prd\circ\tuple{x,\prd\circ\tuple{y,z}}\equiv\prd\circ\tuple{\prd\circ\tuple{x,y},z} $} \\
  } \\
   \cline{2-3}
   \end{array} $$
However, many details are implicit in this illustration.
More precisely, the equational specification $\Ss_{\sgp}$ can be built 
as follows.
First an equational specification $\Ss_{\mgm}$ for magmas 
(a \emph{magma} is simply a set with a binary operation)
is made of 
two types $G$ and $G^2$, three terms $u,v,\prd:G^2\to G$
and one potential product $G\lupto{u}G^2\rupto{v}G$. 
Then, a second equational specification $\Ss_{\mgm}'$ is 
obtained by adding to $\Ss_{\mgm}$
a type $G^3$,
terms $x:G^3\to G$, $w:G^3\to G^2$,
a potential product $G\lupto{x}G^3\rupto{w}G^2$,
and also the terms $f_1=\prd\circ\tuple{x,\prd\circ w}:G^3\to G$
and $f_2=\prd\circ\tuple{\prd\circ \tuple{x,u\circ w},v\circ w}:G^3\to G$.
We also add $y=u\circ w$ and $z=v\circ w$, 
and the equations $w\equiv \tuple{y,z}$,
$f_1\equiv \prd\circ\tuple{x,\prd\circ\tuple{y,z}}$
and $f_2\equiv \prd\circ\tuple{\prd\circ\tuple{x,y},z}$. 
Then $\Ss_{\mgm}'$ is equivalent to $\Ss_{\mgm}$. 
Finally $\Ss_{\sgp}$ is made of $\Ss_{\mgm}'$ with 
the equation $f_1\equiv f_2$, or equivalently with the equation 
$\prd\circ\tuple{x,\prd\circ\tuple{y,z}}\equiv\prd\circ\tuple{\prd\circ\tuple{x,y},z}$.
\end{exam}

In sections~\ref{subsec:dia-spec} and~\ref{subsec:dia-thry}, 
these notions are embedded in the definition of a \emph{diagrammatic equational logic}
$\logL_{\eq}$ \cite{Du07}.
This means that we build 
a limit sketch $\skE_{\eq,S}$ for $\catS_{\eq}$,
a limit sketch $\skE_{\eq,T}$ for $\catT_{\eq}$,
and a morphism of limit sketches $\ske_{\eq}:\skE_{\eq,S}\to\skE_{\eq,T}$
such that the inclusion functor $G_{\eq}:\catT_{\eq}\to\catS_{\eq}$ 
is the precomposition with $\ske_{\eq}$
and its left adjoint $F_{\eq}:\catS_{\eq}\to\catT_{\eq}$ 
(as we saw in section~\ref{subsec:sketch}) 
maps each equational specification to its generated theory.

\subsection{Equational specifications}
\label{subsec:dia-spec}

In this section we provide a detailed construction of a limit sketch $\skE_{\eq,S}$ 
for the category $\catS_{\eq}$ of equational specifications;
except for equations, we will get essentially the same sketch as $\skE_{\spec}$
in section~\ref{subsec:sk-free}. 
We begin with the sketch $\skE_{\gr}$ for graphs: 
$$
  \xymatrix@C=4pc{
    \;\Type\; & \;\Term\; \ar[l]_{\dom} \ar@<1ex>[l]^{\codom} \\
    }
$$
Then, we extend $\skE_{\gr}$ for each kind of potential features;
each limit sketch is followed by its image by its Yoneda contravariant realization.
Finally, by glueing together these extensions of $\skE_{\gr}$ (by a colimit of limit sketches) 
we get the limit sketch $\skE_{\eq,S}$.

\subsubsection*{$\bullet$ Composites}

A sketch $\skE_{\grco}$ for graphs with potential composites
is obtained by extending $\skE_{\gr}$ as follows, with its potential limit and equalities: 
$$
  
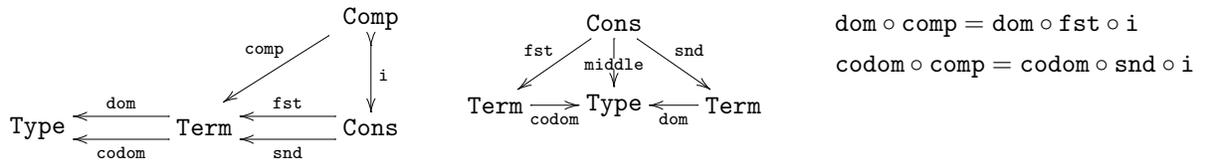
\xymatrix@C=3pc{
    && \hau\Comp \ar@{>->}[d]^{\tti} \ar@<-1ex>[dl]_{\comp} \\
    \Type & \Term \ar@<1ex>[l]^{\codom} \ar@<-1ex>[l]_{\dom} &
    \Cons \ar@<1ex>[l]^{\snd} \ar@<-1ex>[l]_{\fst} \\
    }
\qquad
\xymatrix@=1.5pc{
  & \Cons \ar[dl]_{\fst} \ar[dr]^{\snd} \ar[d]|{\midd} & \\
  \Term \ar[r]_{\codom} & \Type & \Term \ar[l]^{\dom} \\
  } 
\qquad
\begin{array}{l} 
\\ 
\dom\circ\comp = \dom\circ\fst\circ\tti \\
\codom\circ\comp = \codom\circ\snd\circ\tti \\ 
\end{array}
$$
The point $\Comp$ stands for the set of
composable terms, the potential mono $\tti$ for the inclusion, and
the arrow $\comp$ for the composition of composable terms.
The image of $\skE_{\grco}$ by its Yoneda contravariant realization
is the following morphism of realizations of $\skE_{\grco}$;
as required, the image of the mono $\tti$ is an epimorphism.

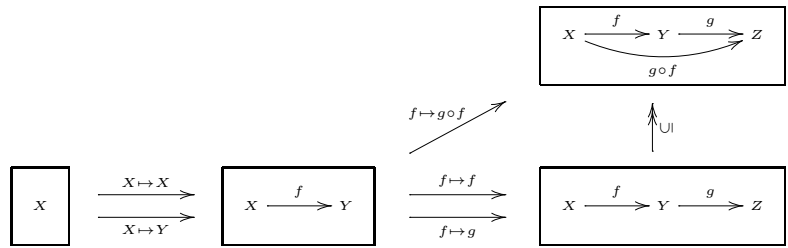
\tiny
$$ \begin{array}{|c|c|c|c|c|}
  \cline{5-5}
  \multicolumn{4}{c|}{ } &
    \xymatrix{
    X \ar[r]^{f} \ar@/_4ex/[rr]_{g\circ f} & Y  \ar[r]^{g} & Z \\}  \\
  \cline{5-5}
  \multicolumn{3}{c}{ } &
  \multicolumn{1}{c}{ \xymatrix@C=3pc@R=1.5pc{
    \mbox{ } & \mbox{ } \\
    \mbox{ } \ar[ru]^{f\mapsto g\circ f} & \mbox{ } \\ } } &
  \multicolumn{1}{c}{  \xymatrix@C=3pc@R=1.5pc{
    \mbox{} \\ \mbox{} \ar@{->>}[u]_{\upseteq}  \\}  } \\
  \cline{1-1}\cline{3-3}\cline{5-5}
  \xymatrix{
    X \\}  &
  \xymatrix@C=3pc{
    \mbox{} \ar@<2ex>[r]^{X\mapsto X}  \ar@<-2ex>[r]_{X\mapsto Y}& \mbox{} \\}  &
  \xymatrix{
    X \ar[r]^{f} & Y \\}  &
  \xymatrix@C=3pc{
    \mbox{} \ar@<2ex>[r]^{f\mapsto f}  \ar@<-2ex>[r]_{f\mapsto g}& \mbox{} \\} &
  \xymatrix{
    X \ar[r]^{f} & Y  \ar[r]^{g} & Z \\} \\
  \cline{1-1}\cline{3-3}\cline{5-5}
  \end{array}$$
\normalsize

\subsubsection*{$\bullet$ Identities}

A sketch $\skE_{\grid}$ for graphs with potential identities
is obtained by extending $\skE_{\gr}$ as follows:
$$
  
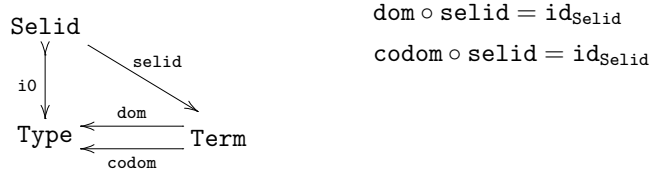
\xymatrix@C=3pc{
    \hau \Selid \ar@{>->}[d]_{\ttiz} \ar@<1ex>[rd]^{\selid} & \\
    \Type & \Term \ar@<1ex>[l]^{\codom} \ar@<-1ex>[l]_{\dom} \\
    }
\qquad\qquad
\begin{array}{l} 
\dom\circ\selid = \ttid_{\Selid} \\
\codom\circ\selid = \ttid_{\Selid} \\ 
\end{array}$$
The point $\Selid$ stands for the set of types with a selected identity,
the potential mono $\ttiz$ for the inclusion,
and the arrow $\selid$ for the selection of the identities.

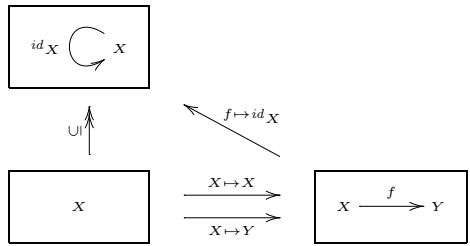
\tiny
$$ \begin{array}{|c|c|c|}
  \cline{1-1}
   \xymatrix{
    \haut X \ar@(lu,ld)_{\id_X} \\}  & \multicolumn{2}{c}{ } \\
  \cline{1-1}
  \multicolumn{1}{c}{  \xymatrix@C=3pc@R=1.5pc{
    \mbox{} \\ \mbox{} \ar@{->>}[u]^{\upseteq}  \\}  } &
  \multicolumn{1}{c}{ \xymatrix@C=3pc@R=1.5pc{
    \mbox{ } & \mbox{ } \\
    \mbox{ } & \mbox{ } \ar[lu]_{f\mapsto\id_X}  \\ } } &
  \multicolumn{1}{c}{ } \\
  \cline{1-1}\cline{3-3}
  \xymatrix{
    X \\ }  &
  \xymatrix@C=3pc{
    \mbox{} \ar@<2ex>[r]^{X\mapsto X}  \ar@<-2ex>[r]_{X\mapsto Y} & \mbox{} \\} &
  \xymatrix{
    X \ar[r]^{f} & Y \\}  \\
  \cline{1-1}\cline{3-3}
  \end{array}$$
\normalsize

\subsubsection*{$\bullet$ Binary products}

A sketch $\skE_{\grpr}$ for graphs with potential binary products
is obtained by extending $\skE_{\gr}$ as follows:
$$
  
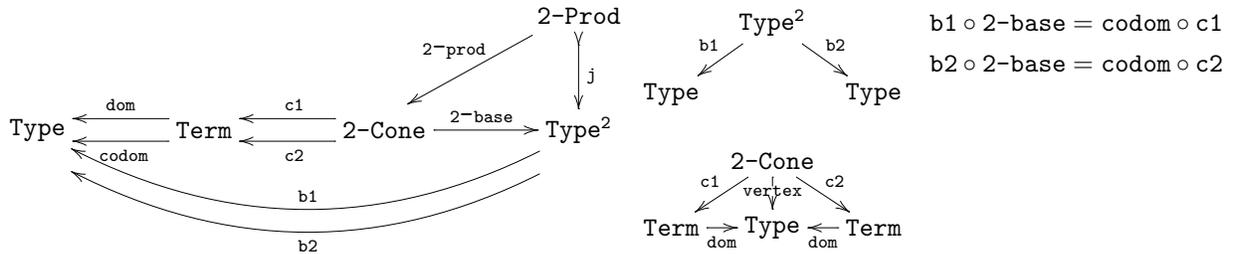
\xymatrix@C=3pc{
    & & & \hau \bProd \ar@{>->}[d]^{\ttj}  \ar@<-1ex>[ld]_{\bprod} \\
    \Type & \Term \ar@<1ex>[l]^{\;\codom} \ar@<-1ex>[l]_{\dom} &
       \bCone  \ar@<1ex>[l]^{\prctwo} \ar@<-1ex>[l]_{\prcone} \ar[r]^{\bbase} &
       \bType  \ar@/^7ex/@<2ex>[lll]^{\prbtwo} \ar@/^7ex/[lll]_{\prbone} \\
    }
 \;
  \xymatrix@R=1pc@C=.5pc{
  & \bType \ar[dl]_{\prbone}  \ar[dr]^{\prbtwo} & \\
  \Type  && \Type \\
  & \bCone \ar[dl]_{\prcone}  \ar[dr]^{\prctwo} \ar[d]|(.45){\vertex}  & \\
  \Term \ar[r]_{\dom} & \Type  & \Term \ar[l]^{\dom} \\
  }
 \;
\begin{array}{l} 
  \\ 
 \prbone\circ\bbase = \codom\circ\prcone \\
  \prbtwo\circ\bbase = \codom\circ\prctwo \\ 
\end{array}$$
The point $\bProd$ stands for the set of binary products, the mono
$\ttj$ for the inclusion, and the arrow $\bprod$ for the
operation which maps a binary product to its underlying binary cone.

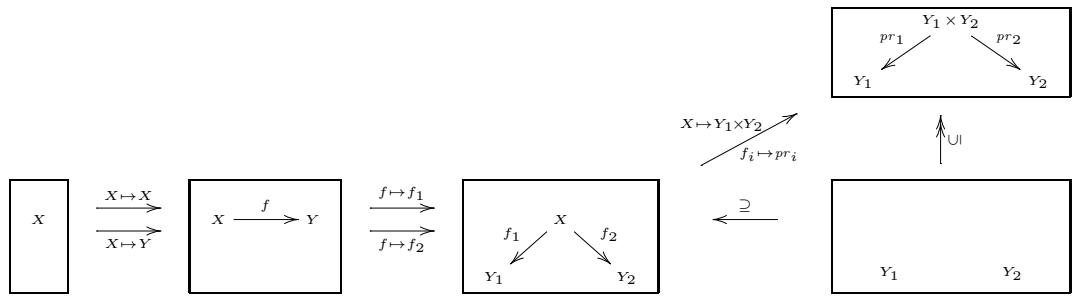
\tiny
$$ \begin{array}{|c|c|c|c|c|c|c|}
  \cline{7-7}
  \multicolumn{6}{c|}{ } &
  \xymatrix@=1pc{
    & Y_1\stimes Y_2 \ar[ld]_{\proj_1} \ar[rd]^{\proj_2}
      & \\
    Y_1 && Y_2  \\}   \\
  \cline{7-7}
  \multicolumn{5}{c}{ } &
  \multicolumn{1}{c}{ \xymatrix@C=3pc@R=1.5pc{
    \mbox{ } & \mbox{ } \\
    \mbox{ } \ar[ru]_{f_i\mapsto \proj_i}^{X\mapsto Y_1\stimes Y_2} & \mbox{ } \\ } } &
  \multicolumn{1}{c}{  \xymatrix@C=3pc@R=1.5pc{
    \mbox{} \\ \mbox{} \ar@{->>}[u]_{\upseteq}  \\}  } \\
  \cline{1-1}\cline{3-3}\cline{5-5}\cline{7-7}
  \xymatrix{
    X \\}  &
  \xymatrix@C=2pc{
    \mbox{} \ar@<2ex>[r]^{X\mapsto X}  \ar@<-2ex>[r]_{X\mapsto Y} & \mbox{} \\}  &
  \xymatrix{
    X \ar[r]^{f} & Y \\}  &
  \xymatrix@C=2pc{
    \mbox{} \ar@<2ex>[r]^{f\mapsto f_1}  \ar@<-2ex>[r]_{f\mapsto f_2} & \mbox{} \\}  &
  \xymatrix@=1pc{
    & X \ar[ld]_{f_1} \ar[rd]^{f_2}  & \\
    Y_1 && Y_2  \\}  &
  \xymatrix@C=2pc{
    \mbox{} & \mbox{} \ar[l]_{\supseteq} \\}  &
  \xymatrix@=1pc{
    & \mbox{ } & \\
    Y_1 && Y_2  \\}   \\
  \cline{1-1}\cline{3-3}\cline{5-5}\cline{7-7}
  \end{array}$$
\normalsize

\subsubsection*{$\bullet$ Pairing}

Now, a sketch $\skE_{\grpair}$ for graphs with potential binary products
and with potential pairings (or 2-tuples) 
is obtained by extending $\skE_{\grpr}$ as follows:
$$
  \xymatrix@C=3pc{
    & & \hau\Pair\ar[r]^{\bcodom}\ar@{>->}[d]_{\bdom}\ar[ld]_{\pair} & 
   \hau \bProd \ar@{>->}[d]^{\ttj}  \ar@<-1ex>[ld]_{\bprod} \\
    \Type & \Term \ar@<1ex>[l]^{\codom} \ar@<-1ex>[l]_{\dom} &
       \bCone  \ar@<1ex>[l]^{\prctwo} \ar@<-1ex>[l]_{\prcone} \ar[r]^{\bbase} &
       \bType  \ar@/^7ex/@<2ex>[lll]^{\prbtwo} \ar@/^7ex/[lll]_{\prbone} \\
    }
\qquad
\begin{array}{l} 
 \ttj\circ\bcodom = \bbase\circ\bdom \\
  \dom\circ\pair = \vertex\circ\bdom \\
  \codom\circ\pair = \vertex\circ\bprod\circ\bcodom \\ 
\end{array}$$

\tiny
$$ \begin{array}{|c|c|c|c|c|c|c|}
  \cline{5-5}\cline{7-7}
  \multicolumn{4}{c|}{ } &
  \xymatrix@=1pc{
    & Y_1\stimes Y_2 \ar[ld]_{\proj_1} \ar[rd]^{\proj_2}
      & \\
    Y_1 && Y_2  \\
    & X \ar[lu]^{f_1} \ar[ru]_{f_2} \ar[uu]^{g}   & \\} &
  \xymatrix@C=2pc{
    \mbox{} & \mbox{} \ar[l]_{\supseteq} \\}  &
  \xymatrix@=1pc{
    & Y_1\stimes Y_2 \ar[ld]_{\proj_1} \ar[rd]^{\proj_2}
      & \\
    Y_1 && Y_2  \\}   \\
  \cline{5-5}\cline{7-7}
  \multicolumn{3}{c}{ } &
  \multicolumn{1}{c}{ \xymatrix@C=3pc@R=1.5pc{
    \mbox{ } & \mbox{ } \\
    \mbox{ } \ar[ru]^{f\mapsto g} & \mbox{ } \\ } } &
  \multicolumn{1}{c}{  \xymatrix@C=3pc@R=1.5pc{
    \mbox{} \\ \mbox{} \ar@{->>}[u]_{\upseteq}  \\}  } &
  \multicolumn{1}{c}{ \xymatrix@C=3pc@R=1.5pc{
    \mbox{ } & \mbox{ } \\
    \mbox{ } \ar[ru]_{f_i\mapsto \proj_i}^{X\mapsto Y_1\stimes Y_2} & \mbox{ } \\ } } &
  \multicolumn{1}{c}{  \xymatrix@C=3pc@R=1.5pc{
    \mbox{} \\ \mbox{} \ar@{->>}[u]_{\upseteq}  \\}  } \\
  \cline{1-1}\cline{3-3}\cline{5-5}\cline{7-7}
  \xymatrix{
    X \\}  &
  \xymatrix@C=2pc{
    \mbox{} \ar@<2ex>[r]^{X\mapsto X}  \ar@<-2ex>[r]_{X\mapsto Y} & \mbox{} \\}  &
  \xymatrix{
    X \ar[r]^{f} & Y \\}  &
  \xymatrix@C=2pc{
    \mbox{} \ar@<2ex>[r]^{f\mapsto f_1}  \ar@<-2ex>[r]_{f\mapsto f_2} & \mbox{} \\}  &
  \xymatrix@=1pc{
    Y_1 && Y_2  \\
    & X \ar[lu]^{f_1} \ar[ru]_{f_2} & \\} &
  \xymatrix@C=2pc{
    \mbox{} & \mbox{} \ar[l]_{\supseteq} \\}  &
  \xymatrix@=1pc{
    Y_1 && Y_2  \\ 
    & \mbox{ } & \\ } \\
  \cline{1-1}\cline{3-3}\cline{5-5}\cline{7-7}
  \end{array}$$
\normalsize

\subsubsection*{$\bullet$ Terminal type}

A terminal (or final) type is a nullary product, and a nullary cone is simply a type (its vertex).
With this correspondence in mind,
the construction of $\skE_{\grfi}$ below is similar to the construction
of $\skE_{\grpr}$ above, with 
$\zCone=\Type$, $\zType=\Unit$ and $\zbase:\Type\to\Unit$, 
where $\Unit$ is the vertex of a potential limit cone with an empty base,
and in addition $\zProd=\Fin$ and $\zprod=\fin$

$$
  \xymatrix@C=3pc{
    \hau \Fin \ar@{>->}[d]_{\ttjz} \ar@<1ex>[rd]^{\fin} & & \\
    \Unit & \Type \ar[l] & \Term \ar@<1ex>[l]^{\codom} \ar@<-1ex>[l]_{\dom} \\
    }
\qquad\qquad
\xymatrix@R=1pc{
  & \Unit & \\
  \mbox{} \ar@{}[rr]|{\txt{(empty base)}} & & \mbox{}  \\
  }
$$

The point $\Unit$ stands for a singleton, 
the point $\Fin$ with the arrow $\ttjz$ for a set with at most one element,
and the arrow $\fin$ stands for the selection of the terminal type.

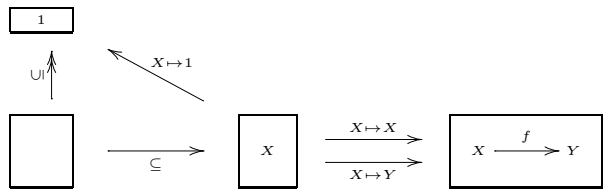
\tiny
$$ \begin{array}{|c|c|c|c|c|}
  \cline{1-1}
   \xymatrix{
    \uno \\}  & \multicolumn{4}{c}{ } \\
  \cline{1-1}
  \multicolumn{1}{c}{  \xymatrix@C=3pc@R=1.5pc{
    \mbox{} \\ \mbox{} \ar@{->>}[u]^{\upseteq}  \\}  } &
  \multicolumn{1}{c}{ \xymatrix@C=3pc@R=1.5pc{
    \mbox{ } & \mbox{ } \\
    \mbox{ } & \mbox{ } \ar[lu]_{X\mapsto \uno}  \\ } } &
  \multicolumn{1}{c}{ } \\
  \cline{1-1}\cline{3-3}\cline{5-5}
  \mbox{ }  &
  \xymatrix@C=3pc{
    \mbox{} \ar[r]_{\subseteq} & \mbox{} \\} &
  \xymatrix{
    X \\ }  &
  \xymatrix@C=3pc{
    \mbox{} \ar@<2ex>[r]^{X\mapsto X}  \ar@<-2ex>[r]_{X\mapsto Y} & \mbox{} \\} &
  \xymatrix{
    X \ar[r]^{f} & Y \\}  \\
  \cline{1-1}\cline{3-3}\cline{5-5}
  \end{array}$$
\normalsize

\subsubsection*{$\bullet$ Collapsing}

Now, a sketch $\skE_{\grcoll}$ for graphs with a potential terminal type 
and with potential collapsings (or 0-tuples) 
is obtained by extending $\skE_{\grfi}$ as follows:
$$
  
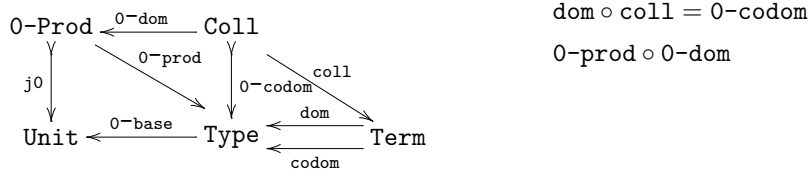
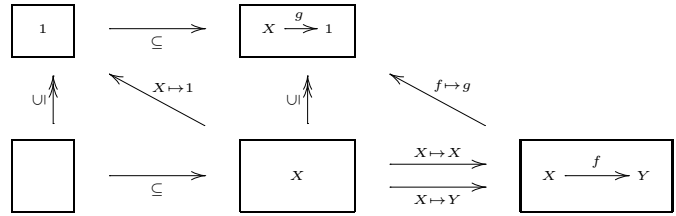
\xymatrix@C=3pc{
    \hau \zProd \ar@{>->}[d]_{\ttjz} \ar@<1ex>[rd]^{\zprod} & 
      \hau\Coll\ar[l]_{\zdom}\ar@{>->}[d]^{\zcodom}\ar[rd]^{\coll} & \\
    \Unit & \Type \ar[l]_{\zbase} & \Term \ar@<1ex>[l]^{\codom} \ar@<-1ex>[l]_{\dom} \\
    }
\qquad\qquad
\begin{array}{l} 
  \dom\circ\coll = \zcodom \\
 \zprod\circ\zdom \\
\end{array}$$

\tiny
$$ \begin{array}{|c|c|c|c|c|}
  \cline{1-1}\cline{3-3}
   \xymatrix{
    \uno \\}  &
   \xymatrix@C=3pc{
    \mbox{} \ar[r]_{\subseteq} & \mbox{} \\}  &
    \xymatrix@=1pc{
    X \ar[r]^{g} & \uno\\}&
    \multicolumn{2}{c}{ } \\
  \cline{1-1}\cline{3-3}
  \multicolumn{1}{c}{  \xymatrix@C=3pc@R=1.5pc{
    \mbox{} \\ \mbox{} \ar@{->>}[u]^{\upseteq}  \\}  } &
  \multicolumn{1}{c}{ \xymatrix@C=3pc@R=1.5pc{
    \mbox{ } & \mbox{ } \\
    \mbox{ } & \mbox{ } \ar[lu]_{X\mapsto \uno}  \\ } } &
   \multicolumn{1}{c}{  \xymatrix@C=3pc@R=1.5pc{
    \mbox{} \\ \mbox{} \ar@{->>}[u]^{\upseteq}  \\}  } &
      \multicolumn{1}{c}{ \xymatrix@C=3pc@R=1.5pc{
    \mbox{ } & \mbox{ } \\
    \mbox{ } & \mbox{ } \ar[lu]_{f\mapsto g}  \\ } }\\
  \cline{1-1}\cline{3-3}\cline{5-5}
  \mbox{ }  &
  \xymatrix@C=3pc{
    \mbox{} \ar[r]_{\subseteq} & \mbox{} \\} &
  \xymatrix{
    X \\ }  &
  \xymatrix@C=3pc{
    \mbox{} \ar@<2ex>[r]^{X\mapsto X}  \ar@<-2ex>[r]_{X\mapsto Y} & \mbox{} \\} &
  \xymatrix{
    X \ar[r]^{f} & Y \\}  \\
  \cline{1-1}\cline{3-3}\cline{5-5}
  \end{array}$$
\normalsize

\subsubsection*{$\bullet$ Equations}

A sketch $\skE_{\greq}$ for graphs with equations is
obtained by extending $\skE_{\gr}$ with two points $\Para$ and
$\Equa$ which stand for the set of pairs of parallel arrows and
the set of equations, respectively. The arrows $\lft$ and $\rght$ extract
the two terms from a pair of parallel terms. 
The potential limit establishes that $\Para$ represents pairs of parallel terms.

$$
  
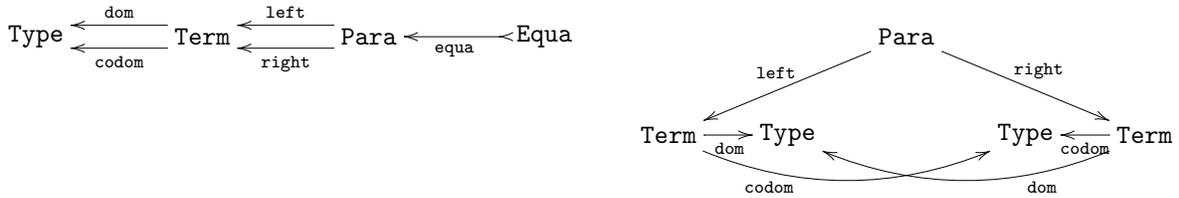
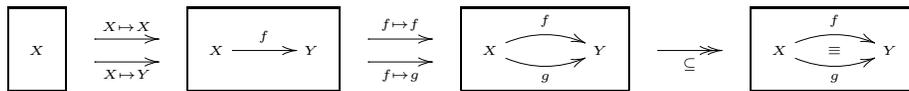
\xymatrix@C=3pc{
    \Type & \Term \ar@<1ex>[l]^{\codom} \ar@<-1ex>[l]_{\dom} &
       \Para  \ar@<1ex>[l]^{\rght} \ar@<-1ex>[l]_{\lft}  &\ \Equa \ar@{>->}[l]^{\equa}  \\
    }
\qquad
\xymatrix@C=1.5pc{
  && \Para \ar[dll]_{\lft} \ar[drr]^{\rght} && \\
  \Term\ar@/_4ex/[rrr]_(.3){\codom}\ar[r]_{\,\dom}& \Type &&
  \Type & \Term \ar@/^4ex/[lll]^(.3){\dom}\ar[l]^{\codom} \\
   \\
  }
$$

\tiny
$$ \begin{array}{|c|c|c|c|c|c|c|}
  \cline{1-1}\cline{3-3}\cline{5-5}\cline{7-7}
  \xymatrix{
    X \\}  &
  \xymatrix@C=2pc{
    \mbox{} \ar@<2ex>[r]^{X\mapsto X}  \ar@<-2ex>[r]_{X\mapsto Y} & \mbox{} \\}  &
  \xymatrix{
    X \ar[r]^{f} & Y \\}  &
  \xymatrix@C=2pc{
    \mbox{} \ar@<2ex>[r]^{f\mapsto f}  \ar@<-2ex>[r]_{f\mapsto g} & \mbox{} \\}  &
  \xymatrix@=1pc{
     X \ar@/^/[rr]^{f}\ar@/_/[rr]_{g}& &Y  \\}  &
  \xymatrix@C=2pc{
    \mbox{} \ar@{->>}[r]_{\subseteq} & \mbox{} \\}  &
  \xymatrix@=1pc{
    X\ar@/^/[rr]^{f}\ar@{}[rr]|{\equiv}\ar@/_/[rr]_{g} & & Y
     \\}   \\
  \cline{1-1}\cline{3-3}\cline{5-5}\cline{7-7}
  \end{array}$$
\normalsize

\subsubsection*{$\bullet$ Equational specifications}

Finally, a sketch $\skE_{\eq,S}$ for equational specifications
is obtained as the colimit of the sketches $\skE_{\grco}$,
$\skE_{\grid}$, $\skE_{\greq}$, $\skE_{\grpair}$ and $\skE_{\grcoll}$
over $\skE_{\gr}$. 
Here is its underlying graph (with $\Type$ repeated twice for readablity), 
in addition it has all the potential limits and all the equalities from the component sketches. 

$$
 
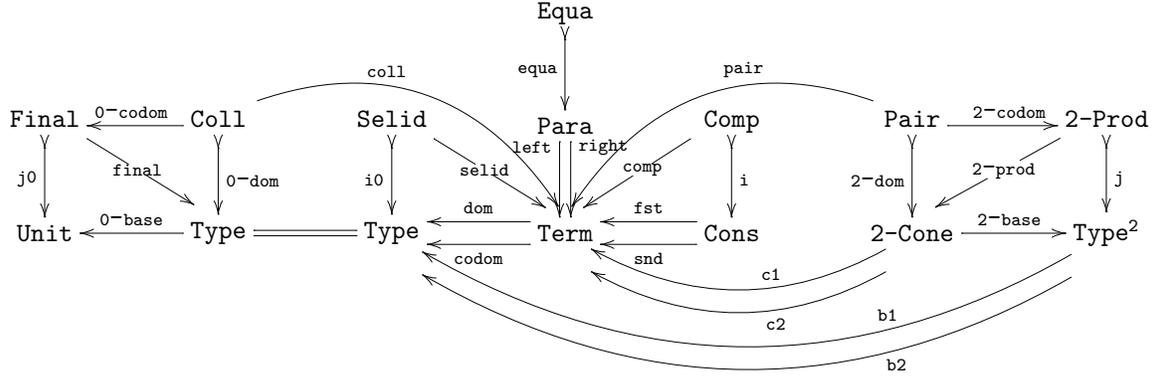
\xymatrix@C=3pc{
   & && \hau \Equa \ar@{>->}[d]_{\equa} &  &  & \\
     \hau \Fin \ar@{>->}[d]_{\ttjz} \ar@<1ex>[rd]|{\fin} & 
         \hau \Coll \ar[l]_{\zcodom} \ar@{>->}[d]^{\zdom}  \ar@/^7ex/@<1ex>[rrd]^(.4){\coll} &
       \hau \Selid \ar@{>->}[d]_{\ttiz} \ar@<1ex>[rd]|{\selid} &
      \Para\ar@<-.5ex>[d]_(.2){\lft}\ar@<.5ex>[d]^(.2){\rght} &
      \hau \Comp \ar@{>->}[d]^{\tti} \ar@<-1ex>[dl]|{\comp}
      & \hau\Pair\ar[r]^{\bcodom}\ar@{>->}[d]_{\bdom}\ar@/_7ex/@<-1ex>[lld]_(.4){\pair}& 
         \hau \bProd \ar@{>->}[d]^{\ttj}  \ar@<-1ex>[ld]|{\bprod} \\
   \Unit & \Type \ar[l]_{\zbase} \ar@{=}[r] & \Type & \Term \ar@<1ex>[l]^{\codom} \ar@<-1ex>[l]_{\dom} &
      \Cons \ar@<1ex>[l]^{\snd} \ar@<-1ex>[l]_{\fst} &
      \bCone  \ar@/^5ex/@<2ex>[ll]^(.4){\prctwo} \ar@/^5ex/[ll]_(.4){\prcone} \ar[r]^{\bbase} &
      \bType \ar@/^10ex/@<2ex>[llll]^(.3){\prbtwo} \ar@/^10ex/[llll]_(.3){\prbone} \\
    }
$$

\begin{exam}
\label{exam:dia-spec-nat}
Let us consider the equational specification $\Ss_{\nat}$:
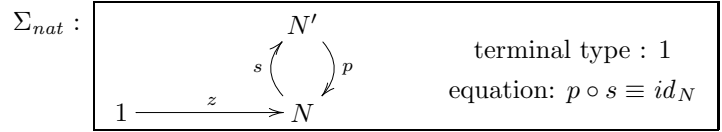
  $$ \begin{array}{c|cc|}
  \cline{2-3}
  \Ss_{\nat} : &
  \xymatrix@R=0.6pc{
    \mbox{} && N' \ar@/^1pc/[dd]^{p} & \\
    \mbox{} && & \\
    \uno \ar[rr]^{z} && N \ar@/^1pc/[uu]^{s} & \\
    } &
  \xymatrix@R=0pc{
  \mbox{ } \\
  \txt{ terminal type : $\uno$} \\
  \txt{ equation: $p\circ s \equiv \id_N$} \\
  } \\
   \cline{2-3}
   \end{array} $$
The specification $\Ss_{\nat}$ has a model ``of naturals'' $M_{\nat}$ which
maps the type $\uno$ to a singleton $\{\star\}$, the type $N$
to the set $\bN$ of non-negative integers, the type $N'$ to the set
$\bN^*$ of positive integers, the term $z$ to the constant function
$\star\mapsto 0$, and the terms $s$ and $p$ to the
functions $x\mapsto x+1$ and $x\mapsto x-1$, respectively.
So, the model $M_{\nat}$ of $\Ss_{\nat}$ is illustrated by a diagram
in the equational theory of sets,
which has the same form as the diagram for $\Ss_{\nat}$:
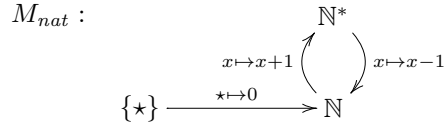
  $$ \begin{array}{cc}
  M_{\nat} : &
  \xymatrix@R=0.6pc{
    \mbox{} && \bN^* \ar@/^1pc/[dd]^{x\mapsto x-1} & \\
    \mbox{} &&& \\
    \{\star\} \ar[rr]^{\star\mapsto0} && \bN \ar@/^1pc/[uu]^{x\mapsto x+1} & \\
    }  \\
   \end{array} \qquad\qquad  $$
Besides, $\Ss_{\nat}$ can be seen as a set-valued realization of $\skE_{\eq,S}$:
\footnotesize
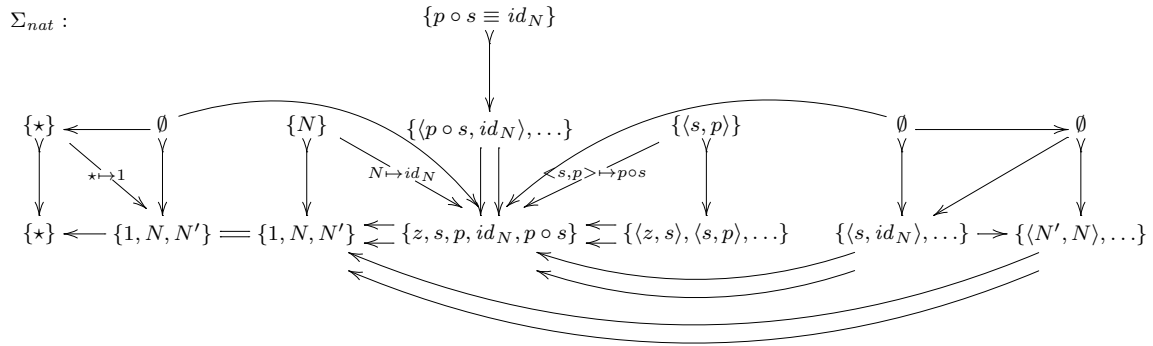
$$
  \xymatrix@C=.9pc{
   \Ss_{\nat}: &  && \hau \{p\circ s \equiv \id_N\} \ar@{>->}[d] &  &  & \\
     \hau \{\star\} \ar@{>->}[d] \ar@<1ex>[rd]|{\star\mapsto\uno} &
     \hau \emptyset \ar[l] \ar@{>->}[d] \ar@/^7ex/@<1ex>[rrd] &     
       \hau \{N\} \ar@{>->}[d] \ar@<1ex>[rd]|{N\mapsto\id_N} &
      \{\tuple{p\circ s, \id_N},\dots\} \ar@<-1ex>[d] \ar@<1ex>[d] &
      \hau \{\tuple{s,p}\} \ar@{>->}[d] \ar@<-1ex>[dl]|{<s,p>\mapsto p\circ s} & 
     \hau \emptyset\ar[r]\ar@{>->}[d]\ar@/_7ex/@<-1ex>[lld] &
       \hau \emptyset \ar@{>->}[d] \ar[ld] \\
   \{\star\} & \{\uno,N,N'\} \ar[l] \ar@{=}[r] &  \{\uno,N,N'\} & 
  \{z,s,p,\id_N,p\circ s\} \ar@<1ex>[l] \ar@<-1ex>[l] &
      \{\tuple{z,s},\tuple{s,p},\dots\} \ar@<1ex>[l] \ar@<-1ex>[l] &
      \{\tuple{s,\id_N},\dots\} \ar@/^5ex/@<2ex>[ll] \ar@/^5ex/[ll] \ar[r] &
      \{\tuple{N',N},\dots\} \ar@/^10ex/@<2ex>[llll] \ar@/^10ex/[llll] \\
    \mbox{ } \\
    }
$$
\normalsize
\end{exam}

\subsection{Equational logic}
\label{subsec:dia-thry}

In section~\ref{subsec:dia-spec}, a limit sketch 
$\skE_{\eq,S}$ for equational specifications has been defined. 
Now, we describe simultaneously a limit sketch $\skE_{\eq,T}$ for equational theories
and a morphism $\ske_{\eq}:\skE_{\eq,S}\to\skE_{\eq,T}$, 
by translating at the sketch level the fact that 
the equational theories are the equational specifications
which satisfy the rules of the equational logic, 
as described in section~\ref{subsec:dia-equ}. 
It happens that this can be done simply by mapping some arrows in $\skE_{\eq,S}$
to identities, thereby some pairs of points in $\skE_{\eq,S}$ get identified in $\skE_{\eq,T}$. 
Let us call \emph{(syntactic) entailment} 
any arrow $t$ in $\skE_{\eq,S}$ which will become an identity in $\skE_{\eq,T}$. 
Let us look more closely at the rules of the equational logic (figure~\ref{fig:equ}). 
Each rule may be considered as a \emph{fraction} 
in the sense of \cite{GZ67}, i.e., as a span $r=\frsyn{s}{t}$ 
from the hypothesis $H$ to the conclusion $C$, 
where the \emph{denominator} $t$ is an entailment,
which is illustrated as follows:
  $$\xymatrix@C=3pc{
     H \ar@<1ex>@{-->}[r] & H'  \ar[l]^{t}  \ar[r]^{s} & C \\
  }$$
The image of a fraction $r=\frsyn{s}{t}$ by the Yoneda contravariant realization 
is a cospan $\rho=\fr{\tau}{\ss}$ in the category $\catS_{\eq}$ of equational specifications,
where $\tau=\funY(t)$, which will become an idenitity in $\catT_{\eq}$,
is also called an \emph{entailment}; this is illustrated in a similar way:
  $$\xymatrix@C=3pc{
     \HH \ar[r]_{\tau} & \HH' \ar@<-1ex>@{-->}[l] & \CC\ar[l]_{\ss}  \\
  }$$
The numerators of the rules are used for easily composing the rules,
but here only the denominators matter,
and in addition it may happen that several rules have the same denominator.
Let $F_{\eq}\dashv G_{\eq}$ be the adjunction associated to $\ske_{\eq}$,
then the functor $F_{\eq}$ is obtained by mapping 
the denominators of the rules to identities.
A similar approach can be found in \cite{Makkai97}.

For instance, the reflexivity rule means that 
``for each term $f$ there is an equation $f\equiv f$'',
the corresponding fraction is: 
  $$\xymatrix@C=3pc{
     \Term \ar@<1ex>@{-->}[r] & \Refl \ar[l]^{t} \ar[r]^{s} & \Equa \\
  }$$
where $\Refl$, $s$ and $t$ are defined by the potential limit:
$$
\xymatrix{
  & \Refl \ar[dl]_{t} \ar[d] \ar[dr]^{s} & \\
  \Term & \Para  \ar@<1ex>[l]^{\rght} \ar@<-1ex>[l]_{\lft} & \Equa \ar[l]^{\equa} \\
  }
$$
The specification $\funY(\Refl)$ is made of a term $f:X\to Y$
and an equation $f\equiv f$, which is represented as 
$ \xymatrix@C=2pc{    X \ar[r]^(.3){f}|{\,\equiv\,} & Y \\ } $.
The image of this rule by Yoneda is:

\tiny
$$ \begin{array}{|c|c|c|c|c|}
  \cline{1-1}\cline{3-3}\cline{5-5}
  \haut  \xymatrix{
    X \ar[r]^{f} & Y \\ } &
  \!\!\!\!\!\! \xymatrix@C=3pc{
    \mbox{} \ar[r]_{f\mapsto f} & \mbox{} \ar@<-1ex>@{-->}[l]  \\}  \!\!\!\!\!\! &
  \xymatrix{
    X \ar[r]^(.3){f}|{\,\equiv\,} & Y \\ } &
  \!\!\!\!\!\! \xymatrix@C=3pc{
    \mbox{} & \mbox{} \ar[l]_{f\mapsto f}^{g\mapsto f} \\}  \!\!\!\!\!\! &
  \xymatrix{
    X \ar@/^/[r]^{f} \ar@/_/[r]_{g} \ar@{}[r]|{\equiv}  & Y \\ } \\
  \cline{1-1}\cline{3-3}\cline{5-5}
  \end{array} $$
\normalsize

In figure~\ref{fig:dia}, for several rules of the equational logic we give 
the corresponding denominator in $\skE_{\eq,S}$ (on the left) 
and its image by $\funY$ in $\catS_{\eq}$ (on the right).
This entailment has the form 
$\xymatrix@C=2pc{\HH \ar[r]_{\tau} & \HH' \ar@<-1ex>@{-->}[l]  \\ }$
and can be read as: 
``as soon as there is an occurrence of $\HH$ in a specification $\Ss$, 
it may be extended (up to equivalence) as an occurrence of $\HH'$''.

\begin{figure}[ht]
\begin{center}
\begin{tabular}{|cccccc|}
\hline
entailment & \cols{5}{image by $\funY$} \\
\hline
\hline
  \cols{6}{composition: ``each pair of consecutive terms is composable''} \\
  $\xymatrix@C=3pc{
     \Cons \ar@<1ex>@{-->}[r] & \Comp \ar[l]^{\tti} \\
  }$ &
  $\quad$ &
\framebox[90pt]{\tiny  $ \xymatrix{
    X \ar[r]^{f} & Y  \ar[r]^{g} & Z \\} $ }&
\tiny \entail &
\framebox[90pt]{\tiny $ \xymatrix{
    X \ar[r]^{f} \ar@/_4ex/[rr]_{g\circ f} & Y  \ar[r]^{g} & Z \\}  $ }&
  $\quad  $ \\[18 pt]
\hline
  \cols{6}{identity: ``each type is a type with identity''} \\
  $\xymatrix@C=3pc{
     \Type \ar@<1ex>@{-->}[r] & \Selid \ar[l]^{\ttiz} \\
  }$ & &
\framebox[90pt]{\tiny $ \xymatrix{
    X \\ } $ }&
\tiny \entail &
\framebox[90pt]{\tiny $   \xymatrix{
    X \ar@(lu,ld)_{\id_X}\ \\}$ }& \\
\hline
  \cols{6}{reflexivity: ``for each term $f$ there is an equation $f\equiv f$''} \\
$\xymatrix@C=3pc{
     \Term \ar@<1ex>@{-->}[r] & \Refl \ar[l]^{t} \\
  }$ & &
\framebox[90pt]{\tiny $ \xymatrix{
    X \ar[r]^{f} & Y \\ }$ }&
\tiny \entail &
\framebox[90pt]{\tiny $ \xymatrix{
    X \ar[r]^(.3){f}|{\,\equiv\,} & Y \\ } $ }& \\
\hline
  \cols{6}{binary product: ``each pair of types has a product''} \\
$\xymatrix@C=3pc{
     \bType \ar@<1ex>@{-->}[r] & \bProd \ar[l]^{\ttj}  \\
  }$ & &
\framebox[90pt]{\tiny $ \xymatrix@=1pc{
    &   & \\
    Y_1 && Y_2  \\} $ } &
\tiny \entailhigh &
\framebox[90pt]{\tiny $ \xymatrix@=1pc{
    & Y_1\times Y_2 \ar[ld]_{\proj_1} \ar[rd]^{\proj_2} \\
    Y_1 && Y_2  \\}  $} & \\[28 pt]
\hline
  \cols{6}{pairing: ``each binary cone has a pairing''} \\
$\xymatrix@C=3pc{
     \bCone \ar@<1ex>@{-->}[r] & \Pair \ar[l]^{\bdom} \\
  }$ & &
\framebox[90pt]{\tiny $ \xymatrix@=1pc{ \\
    Y_1 & & Y_2  \\
    & X \ar[lu]^{f_1} \ar[ru]_{f_2} & \\} $ }&
\tiny \entailhigh &
\framebox[90pt]{\tiny $ \xymatrix@=1pc{
    & Y_1\times Y_2 \ar[ld]_{\proj_1} \ar[rd]^{\proj_2}
      & \\
    Y_1 & \ar@{}[l]|{\equiv} \ar@{}[r]|{\equiv} & Y_2  \\
    & X \ar[lu]^{f_1} \ar[ru]_{f_2} \ar[uu]|{\tuple{f_1,f_2}}   & \\} $} & \\[50 pt]
\hline
  \cols{6}{terminal type: ``there is a terminal type''} \\
$\xymatrix@C=3pc{
     \Unit \ar@<1ex>@{-->}[r] & \zProd \ar[l]^{\ttjz}  \\
  }$ & &
\framebox[90pt]{\tiny $  \xymatrix{\ \\}$}  &
\tiny \entail &
\framebox[90pt]{\tiny $  \xymatrix{\uno \\} $} & \\
\hline
  \cols{6}{collapsing: ``each type is collapsing''} \\
$\xymatrix@C=3pc{
     \Type \ar@<1ex>@{-->}[r] & \Coll \ar[l]^{\zcodom} \\
  }$ & &
\framebox[90pt]{\tiny $ \xymatrix@=1pc{
    X    \\} $} &
\tiny \entail &
\framebox[90pt]{\tiny $ \xymatrix@=1pc{
    X \ar[r]^{g} & \uno\\}$} & \\
\hline
\end{tabular}
\end{center}
\caption{\label{fig:dia} Some rules for equational logic, diagrammatically}
\end{figure}

\end{document}